\documentclass[12pt, letterpaper]{article}
\usepackage[margin=2cm]{geometry}

\usepackage{amssymb}
\usepackage{amsmath}
\usepackage{graphicx}

\usepackage{cite}
\newcommand{\pdv}[2]{\frac{\partial #1}{\partial #2}}

\newcommand{\Li}{\text{Li}}
\usepackage{nicefrac}   
\newcommand{\half}[1]{\nicefrac{#1}{2}}

\usepackage{color}
\newcommand{\change}[1]{#1}

\title{Information geometry for Fermi-Dirac and Bose-Einstein quantum statistics}

\author{Pedro Pessoa$^1$, Carlo Cafaro$^2$ \\
$^1$Department of Physics, University at Albany (SUNY)\\ 
$^2$Department of Mathematics and Physics, SUNY Polytechnic Institute}
\date{}

\begin{document}

\maketitle
\abstract{
Information geometry is an emergent branch of probability theory that consists of assigning a Riemannian  differential geometry structure to the space of probability distributions.
We present an information geometric investigation of gases following the Fermi-Dirac and the Bose-Einstein quantum statistics. For each quantum gas, we study the information geometry of the curved statistical manifolds associated with the grand canonical ensemble.
The Fisher-Rao information metric and the scalar curvature are computed for both fermionic and bosonic models of non-interacting particles. In particular, by taking into account the ground state of the ideal bosonic gas in our information  geometric analysis, we find that the singular behavior of the scalar curvature in the condensation region disappears. This is a counterexample to a long held conjecture that curvature always diverges in phase transitions.
}

\newpage

\section{Introduction}
\paragraph{}

Information geometry (IG) is the application of differential geometry to probability theory \cite{Amari16,Ay17}. 
In this geometric description, the distances are obtained from the Fisher-Rao information metric (FRIM) \cite{Fisher25,Rao45} and represent the distinguishability between neighbouring probability distributions.
This finds application in a large number of information science disciplines including machine learning \cite{Amari10}, signal processing \cite{Wu19} as well as quantum information \cite{Cafaro17} and statistical physics.
More specifically, the role played by geometric techniques in physics cannot be underestimated \cite{Nakahara03}.

Weinhold was one of the first scientists to apply elements of geometry into study of the structure of equilibrium statistical mechanics \cite{Weinhold75} as an extension of Einstein's fluctuation theory and the axioms of thermodynamics. It was Ruppeiner who first introduced a well-defined Riemannian metric structure as the Hessian of the thermodynamical entropy \cite{Ruppeiner79}. 
In a modern understanding, this geometrical structure in statistical physics is due to Jaynes' description of thermodynamics from information theory \cite{Jaynes57,Jaynes57b}, in which Gibbs distributions arise from maximization of entropy (MaxEnt) under expected value constraints. The geometric structure found by Ruppeiner is equivalent to the IG of Gibbs distributions \cite{Brody95,Caticha15}.

Gibbs distributions defined over the space of second quantization in quantum mechanics -- usually referred to as Fock space and parametrized by the number of particles occupying each possible quantum state -- are known in thermodynamics as the Fermi-Dirac (FD) statistics for fermionic particles and the Bose-Einstein (BE) statistics for bosonic particles. These distributions are important not only to provide foundations of statistical mechanics by describing the systems with quantum mechanics, but are also particularly relevant for material sciences, astrophysics and cosmology. 

The Riemannian geometric formulation of thermodynamics and statistical mechanics has opened up a wide range of applications extended to investigations concerning complexity, phase transitions, and critical phenomena \cite{Janke04,Ruppeiner15,Felice18}. A key new tool introduced  is  Riemannian scalar curvature. 
The original result by Ruppeiner \cite{Ruppeiner79} is that the scalar curvature of the manifold  characterizing a classical ideal gas is identically zero.
Expanding on that,  Ruppeiner proposed \cite{Ruppeiner79} the so-called \textquotedblleft\emph{interaction   hypothesis}\textquotedblright\ according to which the scalar curvature in the context of thermodynamics geometry  can be viewed as a measure of interaction in gas systems \cite{Ruppeiner79,Ruppeiner95}. 
Specifically, if interparticle interactions between particles are not present, as in an ideal gas, the space of thermodynamic states is flat. In the presence of interparticle interactions, the space is curved and exhibits nonzero curvature. In particular, curvature is proportional to the (correlation) volume and diverges to infinity near a critical point. 

Despite this very appealing interpretation, as Ruppeiner himself pointed out \cite{Ruppeiner10}, there is no explicit quantitative connection between scalar curvature and phase transitions. 
Also, as argued by Brody and Rivier\cite{Brody95},  there is no definitive physical interpretation of the scalar curvature in the context of information geometric investigations of statistical mechanics of gases. To the best of our knowledge there is no mathematical proof that the scalar curvature diverges at critical points. This divergent behavior remains a conjecture to be proved or disproved by means of a suitable counterexample \cite{Brody95}.
Moreover, the meaning of the sign of the scalar curvature is also unclear. 
To the best of our knowledge, nowadays there is no common agreement on the this sign means \cite{Ruggieri20}. 

In \cite{Janyszek90}, Janyszek and Mrugala  used the concept of quantum mechanical exchange effect to suggest a physical interpretation of the sign of the scalar curvature for ideal quantum gases. 
In the sign convention according to which the scalar curvature of a two-sphere of radius $a$ is positive (that is, $R=2/a^{2}$), it happens that \cite{Janyszek90,Oshima99,Mirza10,Ruggieri20}: i) $R=0$ for an ideal classical gas; ii) $R<0$ for an ideal FD gas; iii) $R>0$ for an ideal BE gas.
Janyszek and Mrugala \cite{Janyszek90} proposed to interpret the scalar curvature as an indicator of stability of the physical system under investigation, with the understanding that a stable system exhibits thermodynamically negligible fluctuations.
Even in a non interactive model, particles in a bosonic gas tend to agglomerate in the same quantum state (that is, attractive exchange effect) leading to a  positive space correlations and, as a consequence, bigger fluctuations. For this reason, bosonic gases are less stable gas than an ideal classical gas and are characterized by a positive $R$. Fermions, instead, are more stable than classical particles since effectively they repeal other particles in the same quantum state (that is, repulsive exchange effect) yielding negative space correlations and, consequently, smaller fluctuations.  Therefore, fermions are specified by a negative $R$.

Posterior studies on the information geometry of quantum gases appear to agree to that interpretation. 
For instance, Oshima and collaborators \cite{Oshima99}   support of instability interpretation proposed by Janyszek and Mrugala  by performing a numerical curvature analysis of an ideal quantum gas obeying Gentile's statistics\footnote{For the sake of completeness, we point out that Gentile statistics \cite{gentile40} is a generalization of the FD and BE quantum statistics where no more than $p$ particles are allowed to occupy the same quantum state. One recovers the FD and the BE statistics when $p=1$ and $p \rightarrow \infty$, respectively.}. 
Similarly, Mirza and Mohammadzade \cite{Mirza09,Mirza10} investigate the IG properties of an ideal gas of fractional statistical particles obeying Gentile's statistics.

Focusing on information geometric investigations of quantum gases \cite{Janyszek90,Oshima99,Mirza10,Quevedo15},
we bring into the discussion that for bosonic systems the number of particles in the ground state is unbounded and it is well-known that at low temperatures (more specifically, in the limit of the fugacity approaching one and the temperature of the gas approaching zero) a thermodynamically relevant fraction of particles reside in the ground state, which explains the phenomena known as BE condensation. Such a phenomena does not appear in FD statistics since there cannot  be more than one particle in the ground state. 
To the best of our knowledge, the information geometric investigations carried out so far into the literature  excluded the BE condensation  from their considerations since these analyses ignored the particles in the ground state \cite{Janyszek90,Oshima99,Mirza10}.

In summary, the present article is motivated by two main reasons. First, the lack of a clear physical interpretation of scalar curvature and its sign motivates us. This lack, in particular, opens up the possibility of proposing theoretical models accommodating critical phenomena where curvature is not necessarily expected to diverge. Second, the lack of an information geometric analysis of an ideal BE gas that includes the BE condensation   drives our  investigation presented here.

Combining information geometry with Jaynes' information theory approach to statistical mechanics, we seek to obtain a general formula for the FRIM for both FD and BE statistical models. In particular, we aim at investigating  geometrical quantities derived from such a metric, including volume elements and scalar curvature. Our approach is unique and original for a number of reasons: (i) Unlike previous works, we calculate the above-mentioned geometrical quantities without assuming the density of states exponent \textquotedblleft$\eta $\textquotedblright\ that specifies the gas model. Therefore, the formulas derived here are applicable to a broader range of quantum systems once $\eta$ is selected; (ii) Unlike previous works, we take into account the particles in the ground state of a bosonic system. To the best of our knowledge, this is the first time IG is done considering the ground state correction for BE statistics. Therefore, our information geometric investigation will give us an appropriate description near BE condensation. For a more transparent interpretation of our results, we also present illustrative plots of suitable geometrical quantities with the help of mpmath python library \cite{mpmath}. In particular, this library gives a precise calculation for the polylogarithm function \cite{wolframpages} which is fundamental for both BE and FD quantum statistics. 
%\change{The codes for all the numerical calculation performed here is available at our GitHub repository \cite{Repository}. }

Is also of relevance  that in our investigation FD and BE statistics provide a quantum description without relying on the density matrices formalism. Therefore, although we calculate FRIM on a model that describes quantum states, this geometric analysis refers directly to probability distributions in the Fock space and it is not an example of quantum information geometry.  We apply information geometry for probabilities on a discrete set of states.
The concept of scalar curvature employed here happens to be rather useful in a number of geometric investigations of quantum mechanical phenomena. For instance, in the theoretical framework developed by Nielsen et al. \cite{Nielsen06}, quantum computing can be viewed as a free fall in a curved geometry: finding the optimal circuit is equivalent to finding the shortest path between two points in an appropriate curved geometry. In this context, curvature analyses on the manifold of the $\mathrm{SU}\left( 2^{n}\right) $ group of $n$-qubit unitary operators with unit determinant are carried out in order to help characterizing global characteristics of geodesic paths and to determine minimal complexity quantum circuits \cite{Brandt10}. 
Moreover, in the framework of quantum critical phenomena, investigations based upon the use of the scalar curvature on the manifold of coupling constants that parametrize a quantum Hamiltonian are performed to study quantum phase transitions \cite{Zanardi07}. In this specific context, the basic idea is that, since distance between
quantum states encodes their degree of distinguishability in a quantitative
manner, crossing a critical point that separates regions with structurally
different phases should give rise to a singular behavior of the metric (and,
possibly, other quantities built from it such as the scalar curvature).

The layout of the rest of the paper is as follows:
In Section II, we will review some known results that are important for the development of this paper, mainly (i) how Gibbs distributions arise from MaxEnt (ii) how to obtain the FRIM and other geometrical quantities for Gibbs distributions and (iii) how the Gibbs distributions of the grand canonical ensemble are defined in Fock spaces.
In Section III, we will discuss the assumptions involved in the continuous approximation, focusing on how different quantum mechanics microscopic models lead to different values for the density of states exponent $\eta$ and how the continuous approximation leads to the polylogarithm family of functions.
In Section IV, we will calculate FRIM, volume elements and scalar curvature for an FD ideal gas and give a graphical presentation of our results.
In Section V, we calculate the same geometrical quantities and graphs for a BE ideal gas and compare them to the ones calculated without the ground state corrections.
In Section VI, we comment on the classical limit of geometrical quantities. Since both low fugacity and high temperature are required for an ideal FD or BE gas to assimilate to a classical ideal gas, curvature can give more about the system's structure not available in the Legendre structure.

%\newpage
\section{Background}
\subsection{MaxEnt and Gibbs distributions}
\paragraph{}

The information theory description of statistical physics \cite{Jaynes57,Jaynes57b} relies on MaxEnt to assign probability distributions $\rho(x)$, defined over a  general physical space $x\in \mathcal{X}$, which maximizes the entropy $S[\rho]$ -- given by the negative Kullback-Leibler divergence \cite{Kullback51} --  under the expected value constraints for a set of $n$ real functions $a_\mu(x) $, that is

\begin{equation} \label{maxent}
\begin{split}
\max_\rho \hspace{.5cm}  &  S[\rho] = - \int dx \ \rho(x) \log \left( \frac{\rho(x)}{q(x)} \right) \ , \\
s.t. \hspace{.5cm} &  \int  dx \ \rho(x) = 1 \\
& %\langle a_i(x) \rangle=
\langle a_\mu(x) \rangle \doteq \int  dx \ \rho(x) a_\mu(x) = A_\mu \ .
\end{split}
\end{equation}
Here we use $\int dx$ which is the appropriate measure of space of states $\mathcal{X}$, in a discrete space (as we will describe soon) $\int dx$ takes the form of a summation for all possible states,  $\sum_x$.

Particular work has to be done in selecting the appropriate space $\mathcal{X}$ where the probabilities were defined and the appropriate prior $q(x)$ (see \cite{Jaynes65}). 
Also the set of expected values $A = \{A_1,A_2, \ldots,A_n\}$ can be identified into statistical physics as quantities such as total energy, total volume, quantity of particles and magnetization.
Note that $S[\rho]$ reduces to Shannon \cite{Shannon48} or Gibbs entropy  when  $q(x)$ is uniform -- discussion on the role of uniform priors in statistical physics can be found in \cite{Caticha12}.  
The distribution that achieves the entropy maximum in \eqref{maxent} is the Gibbs distribution

\begin{equation}\label{canonicaldefinition}
    \rho(x|\lambda) = \frac{q(x)}{Z(\lambda)} \exp \left(-\sum_{\mu=1}^n \lambda^\mu a_\mu(x)\right) \ ,
\end{equation}
where $\lambda = \{\lambda^1,\lambda^2,\ldots,\lambda^n\}$ is a set of Lagrange multipliers (LM) related to the  expected value constraints. $Z(\lambda)$ is the partition function, a normalization factor independent of $x$, obtained as

\begin{equation} \label{partitition}
    Z(\lambda) = \int dx \  q(x) \exp \left(-  \lambda^\mu a_\mu(x) \ \right)  \ .
\end{equation}
For convenience that will become clearer when discussing the geometrical properties, in \eqref{partitition} and in the remainder of this article we use the Einstein summation notation,
$\lambda^\mu A_\mu =  \sum_{\mu=1}^n  \lambda^\mu A_\mu$. 
The expected values can be recovered from the LM in Gibbs distributions by the following identity

\begin{equation} \label{Averages0}
    A_\mu = \pdv{F(\lambda)}{\lambda^\mu} \quad \text{where} \quad F(\lambda) = - \log Z(\lambda) \ .
\end{equation}
In thermodynamics $F$ is identified as the free energy. 

The generality of MaxEnt implies that Gibbs distributions, also known as canonical distributions or the exponential family, are the only distributions for which sufficient statistics exist.  Meaning, functions for which all properties of the distributions can be extracted -- see literature on the Fisher-Darmois-Koopman-Pitman theorem \cite{Daum86} and texts showing how well studied probability distributions can be written in the exponential form \cite{Nielsen09}. 
Also, the functions $a_\mu(x)$, for each of which the expected value constraints were taken in \eqref{maxent}, are the sufficient statistics.

The entropy \eqref{maxent} then can be computed, at its maximum \eqref{canonicaldefinition}, as a function -- rather than a functional -- of the expected values, $S(A) \doteq \max_\rho (S[\rho])$ which is

\begin{equation}\label{thermalentropy}
S(A) = %S[\rho(x|\lambda(A))] = 
- \int dx \ \rho(x|\lambda(A)) \log\frac{ \rho(x|\lambda(A))}{q(x)}=  \lambda^\mu(A) A_\mu - F (\lambda(A)) \ .
\end{equation}
That means  $S(A)$  is the Legendre transformation of $F(\lambda)$ while $A_\mu$ and $\lambda^\mu$ are each others' dual in the Legendre formalism, this implies

\begin{equation}\label{inverselambdaA}
\lambda^{\mu}=\pdv{
S}{ A_\mu} \ .
\end{equation}  
Having assigned the probability distributions of interest and their relationship to information theory we can move onto a geometrical description.

\subsection{Information Geometry}\label{IG}
\paragraph{}
We will present results from information geometry that are particularly useful for thermodynamics and MaxEnt. This will be obtained by focusing on the geometrical properties of Gibbs distributions.
In this description the set of expected values $\lambda$ will be our primary coordinates -- which justifies why they were written with upper indexes $\lambda^\mu$ representative of contravariant objects, later we show how this is equivalent to a description using the expected values $A$ as primary coordinates -- each point in these coordinates define an unique Gibbs distribution $\rho(x|\lambda)$. The metric tensor for this  statistical is defined by FRIM
                                                         
\begin{equation}
\label{firstmetric}
    g_{\mu\nu} = \int dx \ \rho(x|\lambda) \frac{\partial\log \rho(x|\lambda)}{\partial \lambda^\mu}\frac{\partial\log \rho(x|\lambda)}{\partial \lambda^\nu}  .
\end{equation}
The distances obtained from $dl^2 = {g_{\mu\nu} d\lambda^\mu d\lambda^\nu}$ are, up  to a unit defining constant, a measure of distinguishability between two neighbouring distributions $\rho(x|\lambda) $ and  $\rho(x|\lambda+d\lambda)$.
The choice of metric \eqref{firstmetric} is not arbitrary, details on the derivation and properties  can be seen in \cite{Amari16,Caticha15}. For the current discussion it is sufficient to say that FRIM is the only metric properly tailored to preserve desired properties of probability distributions, namely being invariant under Markov embedding \cite{Cencov81,Campbell86}.

Here we are going to focus on useful forms to express FRIM \eqref{firstmetric} for Gibbs distributions. In \eqref{canonicaldefinition} Gibbs distribution were defined in terms of the LMs $\lambda$ however, as they come from \eqref{maxent} they could also be parametrized in terms the expected values $A$ and the two set of parameters are related by \eqref{inverselambdaA}.  This is not more than a change of variables that can be characterized using

\begin{equation} \label{partial A}
    \frac{\partial A_\mu}{\partial\lambda^\nu} = 
     \frac{\partial^2 F}{\partial\lambda^\nu \partial \lambda^\mu} = %- 
    A_\mu A_\nu - \langle a_\mu  a_\nu \rangle \ ,
\end{equation}
where we identify the covariance tensor and its inverse as 
\begin{equation} \label{corrmetric}
    C_{\mu\nu} =  \langle a_\mu a_\nu  \rangle  - A_\mu A_\nu  =- \frac{\partial A_\mu}{\partial\lambda^\nu} \ \ \ ,   \ \ \   C^{\mu\nu} =  - \frac{\partial\lambda^\mu}{\partial A_\nu} \ .
\end{equation}

We present some useful well-known results on the geometrical description of Gibbs distributions, a proof for these identities can be found in Appendix \ref{IGProofs}.

\emph{Identity 1 --} The metric tensor for the manifold of Gibbs distributions is the  covariance matrix, 

\begin{equation} \label{gequalsC}
    g_{\mu\nu} = C_{\mu\nu} \ .
\end{equation}

\emph{Identity 2 --} The covectors for an infinitesimal change in the expected value coordinates are opposite to the infinitesimal change in the Lagrange multipliers

\begin{equation}
    d\lambda_\mu = g_{\mu\nu} d\lambda^\nu = - dA_\mu \ .
\end{equation}
This means the maximization of entropy makes them dual in the Legendre transform \eqref{thermalentropy} and also information geometry makes them dual coordinates of the statistical manifold, which justifies the use of lower indexes (covariant) notation for $A$.

\emph{Identity 3 --}  In the coordinates defined by the LM -- dual to expected values --  the metric and inverse metric are, respectively, the Hessian of  free energy and entropy,

\begin{equation} \label{metricderiv}
    g_{\mu\nu} = -\frac{\partial^2 F }{\partial \lambda^\mu \partial \lambda^\nu}  \quad \text{and} \quad  g^{\mu\nu} =  -\frac{\partial^2 S }{\partial A_\mu \partial A_\nu} \ .
\end{equation}
 
\emph{Identity 4 --}  For two dimensional Gibbs statistical manifolds, the scalar curvature can be written as:

 \begin{equation} \label{curvature2d}
     R = - \frac{1}{2g^2} \det \left[ \begin{matrix}
g_{11} & g_{12} & g_{22} \\ 
\partial_1 g_{11} & \partial_1 g_{12} & \partial_1 g_{22} \\ 
\partial_2 g_{11} & \partial_2 g_{12} & \partial_2 g_{22} \\ 
\end{matrix} \right]
 \end{equation}
 where $g = \det g_{\mu\nu}$ and $\partial_\sigma g_{\mu\nu} = \pdv{}{\lambda^\sigma} g_{\mu\nu}$.
 
Since we have a way to obtain FRIM from the Gibbs distribution, we can complete our background section with the MaxEnt description of quantum gases.

\subsection{MaxEnt in Fock Spaces}
\paragraph{}
MaxEnt requires the choice of appropriate space $\mathcal{X}$ and prior $q(x)$. For the accurate description of quantum gases, the probabilities should be assigned to the space of occupation numbers of each quantum state (Fock space). That means $x = \{x_i\}$ where $i$ corresponds to an enumeration of the eigenstates obtained from quantum mechanics, each corresponds to an eigenvalue of energy $\epsilon_i$. Each quantity $x_i$ takes the value of the number of particles on the $i$th state.
Fermi-Dirac (FD) statistics refers to a state in which  $x_i$  takes binary values $\{0,1\}$, corresponding to the Pauli exclusion principle for fermions, while in Bose-Einstein (BE) statistics $x_i$ can take natural number values $\{0,1, \ldots \}$.
These quantum statistics are usually represented in the grand canonical ensemble, that means the sufficient statistics are chosen as the energy $a_1(x) = \sum_i \epsilon_i x_i $ and the total number of particles $a_2(x) = \sum_i  x_i $.

That leads to a Gibbs distribution \eqref{canonicaldefinition} of the form

\begin{equation}
    \rho(x|\lambda)= \rho(x_1,x_2, \ldots |\lambda)  = \frac{1}{Z(\lambda)} \prod_i e^{-\lambda^1  \epsilon_i x_i} \ e^{ -\lambda^2 x_i} \ ,
\end{equation}
where

\begin{equation}
    Z(\lambda) = \sum_x \prod_i e^{-\lambda^1  \epsilon_i x_i} \ e^{ -\lambda^2 x_i} \ .
\end{equation}
For FD and BE statistics $Z(\lambda)$ is given by

\begin{equation}
    Z_{FD}(\lambda) = \prod_i \left( 1 +   e^{-\lambda^1  \epsilon_i } \ e^{ -\lambda^2} \right)    \quad \text{and} \quad
    Z_{BE}(\lambda) = \prod_i \left( 1 -  e^{-\lambda^1  \epsilon_i } \ e^{ -\lambda^2} \right)^{-1}  \ ,
\end{equation}
respectively. The partition function \eqref{partitition} and free energy \eqref{Averages0} can be written as

\begin{equation} \label{zandf}
    Z_\pm(\lambda) = \prod_i \left( 1 \pm   e^{-\lambda^1  \epsilon_i } \ e^{ -\lambda^2} \right)^{\pm 1 } \quad \text{and} \quad F_\pm(\lambda) = \pm  \sum_i \log\left( 1 \pm   e^{-\lambda^1  \epsilon_i } \ e^{ -\lambda^2} \right) \ ,
\end{equation}
respectively. The upper (or lower) sign in \eqref{zandf} refers to the FD (or BE) statistics. The LM in \eqref{zandf} can be identified in terms of the usual thermodynamical parameters, temperature $T$ and chemical potential $\mu$,

\begin{equation} \label{lambdaTmubeta}
    \lambda^1 = \frac{1}{kT}  \doteq \beta \quad \text{and} \quad \lambda^2 = - \frac{\mu}{kT} \ ,
\end{equation}
where $k$ is the  Boltzmann constant. We will keep expressing geometrical quantities in terms of the LM as they are the appropriate coordinates as described before.

The  succeeding step into the MaxEnt description is to write the expected values in terms of the LM as in \eqref{Averages0}. This yields,

\begin{equation}\begin{split} 
    U_\pm &= A_1{}_\pm = \pdv{F_\pm}{\lambda^1} = \sum_i \frac{\epsilon_i}{   e^{\lambda^1  \epsilon_i } \ e^{ \lambda^2} \pm  1} \\
    N_\pm &= A_2{}_\pm = \pdv{F_\pm}{\lambda^2}= \sum_i \frac{1}{ e^{\lambda^1  \epsilon_i } \ e^{ \lambda^2} \pm 1 }\ . \label{Averages}
\end{split}\end{equation}
If one correctly assigns the spectrum of energies $\epsilon_i$, the statistical physics description of the FD and BE gases would be complete. However to write the summations in \eqref{Averages} in a closed form is not feasible, to the best of our knowledge. The study of quantum gases is usually tackled through approximations in which the summations are replaced by integrals in energy, as we will discuss in the following section.

\section{Continuous approximation}
\paragraph{}
Physical examples however do not rely on summation such as the ones in \eqref{Averages}. Instead quantum gases  and other physically relevant models based on FD and BE statistics are usually treated as a continuous approximation in terms of states energy,

\begin{equation} \label{Transform}
\sum_i % \rightarrow \sum_\epsilon  \mathsf G(\epsilon)
\rightarrow \int d\epsilon \ \mathsf G(\epsilon) \ ,
\end{equation}
where $\mathsf G(\epsilon)$ is the density of states with energy $\epsilon$. 
This is usually justified under the idea that the internal energy of the system, $U$ is much bigger than the differences in energy of the quantum eigenstates, $\max_{ij}|\epsilon_i-\epsilon_j|\ll U$.

In order to better understand the continuous approximation \eqref{Transform} we will provide a first basic example. For the model of a particle in a three dimensional cubic box of edge length L,  the quantum state will be defined by a triplet of integers, $ i = (n_x,n_y,n_z)$ and the energy state will be given by 

\begin{equation} \label{spectrumIG}
\epsilon = \epsilon_i =  \frac{(2\pi \hbar)^2}{2mL^2}   (n_x^2+n_y^2+n_z^2) \ ,
\end{equation}
where $m$ is the mass of the particle. Therefore the sum over states becomes

\begin{equation} \label{TransformIG}
\sum_i = \sum_{n_x=-\infty}^\infty \sum_{n_y=-\infty}^\infty \sum_{n_z=-\infty}^\infty \rightarrow  \mathsf g_s \int_{-\infty}^\infty dn_x \int_{-\infty}^\infty dn_y\int_{-\infty}^\infty dn_z \ ,
\end{equation}
where $\mathsf g_s $ is the number of different possible polarizations given the particle spin $s$, $\mathsf g_s = 2s+1$. 
The triple integral in \eqref{TransformIG} can be evaluated using \eqref{spectrumIG} under spherical coordinates

\begin{equation}
\label{densityIG}
 \sum_i \rightarrow %\int_{-\infty}^\infty dn_x \int_{-\infty}^\infty dn_y\int_{-\infty}^\infty dn_z  = 
 \int_0^\infty d\epsilon \left[ \frac{\mathsf g_s V}{4\pi^2}\left( \frac{2m}{\hbar^2}\right)^{3/2}\right] \epsilon ^{1/2}\ ,
\end{equation}
where $V =L^3$ is the volume of the box.

\begin{table}%[h!]
  \begin{center}
    \begin{tabular}{l|c c |c c}
      \textbf{System} &\multicolumn{2}{c|}{ $3$ dimensions} & \multicolumn{2}{c}{ $D$ dimensions} \\  \hline    
      & $\eta$ & $\kappa$ & $\eta$ & $\kappa$ \\
      \hline 

      Particle in a box & $\half 1$ & $\frac{\mathsf g_s V}{4\pi^2}\left( \frac{2m}{\hbar^2}\right)^{\half{3}}$ & $\half{D}-1$ & $\frac{\mathsf g_s V}{\Gamma(\half D)}\left( \frac{m}{2 \pi \hbar^2}\right)^{\half{D}}$ \\

      Ultrarelativistic particle & $2$  & $\frac{\mathsf g_s V}{2\pi^2} \left(\frac{1}{\hbar c}\right)^3$  & $D-1$ & $\frac{2 \mathsf g_s V}{\Gamma(\half D)} \left(\frac{1}{ 2\sqrt{\pi}\hbar c}\right)^D$ \\
      Harmonic trapped gas & 2 & $ \frac{ \mathsf g_s}{2} \left(\frac{1}{\hbar \omega}\right)^3$ & $D-1$ & $ \frac{  \mathsf g_s}{\Gamma(D)} \left(\frac{1}{\hbar \omega}\right)^D$ \\
    \end{tabular}
  \end{center}
  \caption{Parameters for density of states $\mathsf G (\epsilon)$. The parameter $\kappa$ depends on $ m$, the mass of the particle, and for the harmonic trapped gas depends on the frequency of the harmonic oscillator  $\omega$.}
    \label{tab:tablequantum}
\end{table}

We will proceed with the calculation of the FRIM on models for which the density of states is proportional to a power of the energy, 

\begin{equation}\label{density}
    \mathsf G (\epsilon) = \kappa \epsilon^\eta \ ,
\end{equation}
where the density of states prefactor ($\kappa$) and the density-energy exponent ($\eta$) are constants defined by the quantum energy spectrum. As an example, for a three dimensional particle in the box we can see from \eqref{densityIG} that  $\kappa= \left[ \frac{\mathsf g_s V}{4\pi^2}\left( \frac{2m}{\hbar^2}\right)^{\half{3}}\right]$ and $\eta=\half{1}$. Other examples of quantum systems for which \eqref{density} is valid and their respective values of $\kappa$ and $\eta$ are presented in Table \ref{tab:tablequantum}. Note that per \eqref{density} and \eqref{Transform} $\kappa$ has units of $[\epsilon]^{-(\eta+1)}$, meaning that $\kappa^{-\frac{1}{\eta+1}}$ can serve as a emergent unit of energy defined by the system.
The continuous approximation \eqref{Transform} transforms the average values \eqref{Averages} as

\begin{equation}\begin{split}
    U_{\pm} &= A_\pm^1 = \kappa \int d\epsilon \ \frac{\epsilon^{\eta+1}}{e^{\lambda^1 \epsilon }\ e^{\lambda^2}\pm1}  \\
    N_{\pm} &= A_\pm^2 =  \kappa  \int d\epsilon \  \frac{\epsilon^{\eta}}{e^{\lambda^1 \epsilon }\ e^{\lambda^2}\pm1}  \ . \label{Averages3}
\end{split}\end{equation}

In order to proceed with calculation of these expected values and geometric quantities derived from them, it is useful to identify the polylogarithm function \cite{wolframpages}

\begin{equation} \label{Polylogdef}
\Li(y,\varphi) = \frac{1}{\Gamma(\varphi)} \int_0^\infty du \ \frac{u^{\varphi-1}}{y^{-1} e^u-1} = \sum_{k=1}^\infty \frac{y^k}{k^\varphi}\ ,
\end{equation}
where $\Gamma(\varphi)$ is the Euler's gamma function. A property of the polylogarithm which is useful for obtaining  geometrical quantities is

\begin{equation} \label{Polylogderivative}
\frac{d}{dy} \Li(y,\varphi)   = \frac{1}{y} \Li(y,\varphi-1) \ . 
\end{equation}
The expected values \eqref{Averages3} can be expressed in terms of polylogarithms. Under the appropriate change of variables,  $u=\beta \epsilon$, and defining

\begin{equation} \label{xianddev}
    \xi \doteq e^{-\lambda^2} %= e^{\frac{\mu}{kT}}\ ,
    \ , \quad \frac{d \xi}{d \lambda^2} = -\xi \ ,
\end{equation}
the variable $\xi$ is referred in thermodynamics as fugacity. For \eqref{Averages3} to converge it imples that for FD statistics $0<\xi<\infty$ and for BE statistics $0<\xi<1$.

Substituting \eqref{Polylogdef} into \eqref{Averages3} we obtain

    \begin{equation}\begin{split} \label{Averages4}
    U_\pm = A_1 {}_\pm = \pdv{F}{\lambda^1} &= \mp \ \kappa \frac{ \Gamma(\eta+2)}{\beta^{\eta+2}} \Li(\mp \xi,\eta+2) \\
    N_\pm = A_2 {}_\pm = \pdv{F}{\lambda^2} &= \mp \ \kappa \frac{ \Gamma(\eta+1)}{\beta^{\eta+1}} \Li(\mp \xi,\eta+1) \ .
    \end{split}\end{equation}
Here we expressed the thermodynamical quantities in terms of the the unit corrected inverse temperature, $\beta$ defined in \eqref{lambdaTmubeta}, and the fugacity, $\xi$ defined in \eqref{xianddev}. 
This choice is made because we commit to using the LMs  $\lambda$ as coordinates. Moreover, we can calculate the metric terms using \eqref{metricderiv}, because of this we will not keep track of the covariant structure in the remainder of this article. If one uses a different set of coordinates they would need to express their coordinates in terms of $\beta$ and $\xi$ -- as we do in \eqref{lambdaTmubeta} and \eqref{xianddev} -- and also do the appropriate change of coordinates from \eqref{metricderiv}.
A similar calculation for the free energy \eqref{zandf} in the continuous approximation yields
\begin{equation}\label{Fcon}
    F_\pm(\lambda) = \pm \ \kappa \frac{ \Gamma(\eta+1)}{\beta^{\eta+1}} \Li(\mp \xi,\eta+2) \ .
\end{equation}

Having general expressions for the expected values \eqref{Averages3} allows for the calculation of FRIM for both FD and BE. Although we focus on a unified treatment for both cases, other considerations arising from the continuous approximation suggest that their treatment should be separated. Mainly the fact that \eqref{density} 
assigns no states for $\epsilon = 0$ leads to a fundamental difference in the description of bosons and fermions. The two following sections will calculate the geometrical quantities for the FD and BE gas, respectively.

\section{Information geometry - Fermions}

\paragraph{}
Applying \eqref{Averages4} and \eqref{Fcon} into \eqref{metricderiv} we can calculate the metric terms for FD models, 

\begin{figure}%[h]
  \begin{minipage}[b]{0.45\textwidth}
    \includegraphics[width=\textwidth]{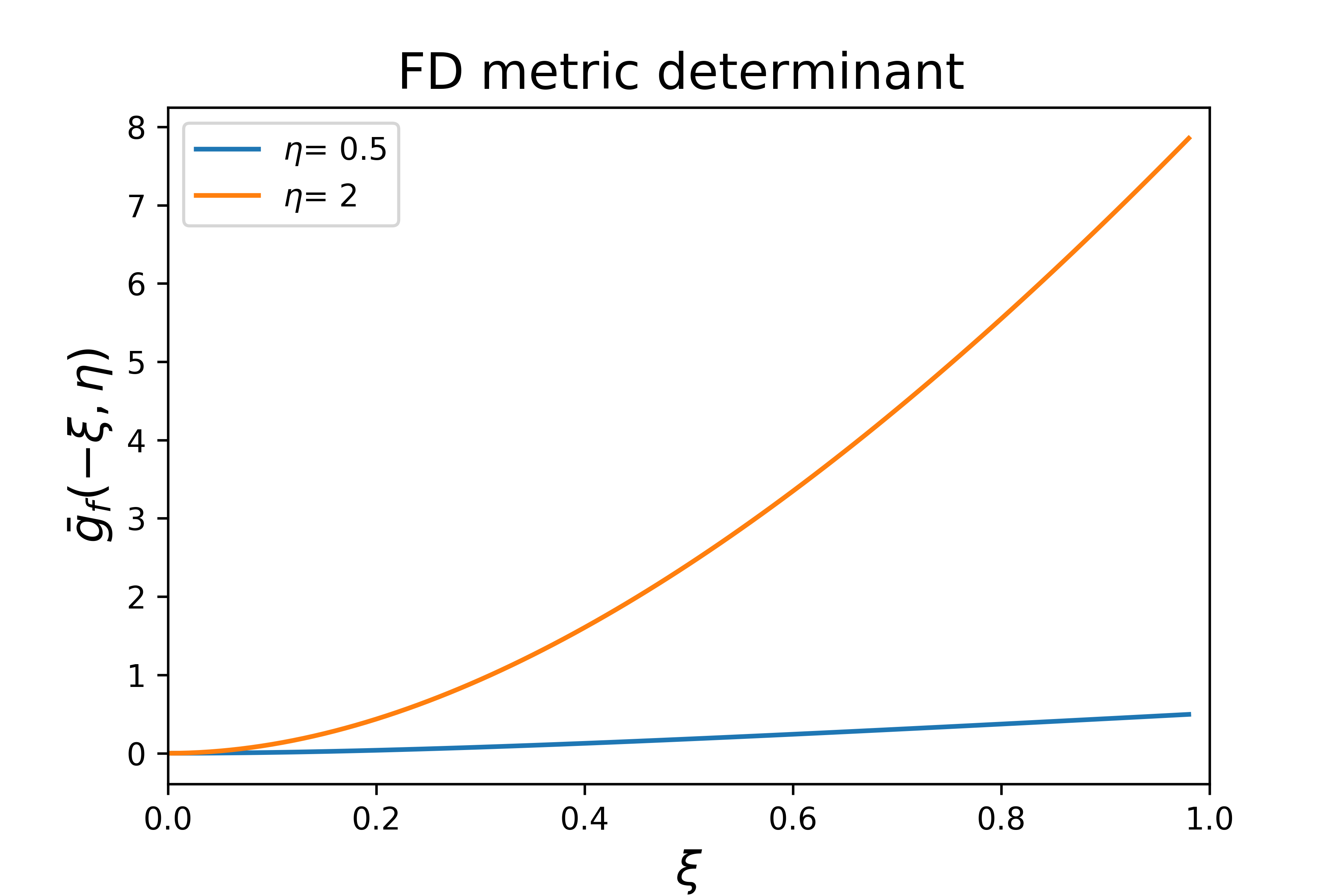}
    \caption{Dimensionless quantity $ \bar{g}_f $, related to metric determinant as \eqref{metricFD} for fermions.}
    \label{fig:Afermions}
  \end{minipage}
  \hfill
  \begin{minipage}[b]{0.45\textwidth}
    \includegraphics[width=\textwidth]{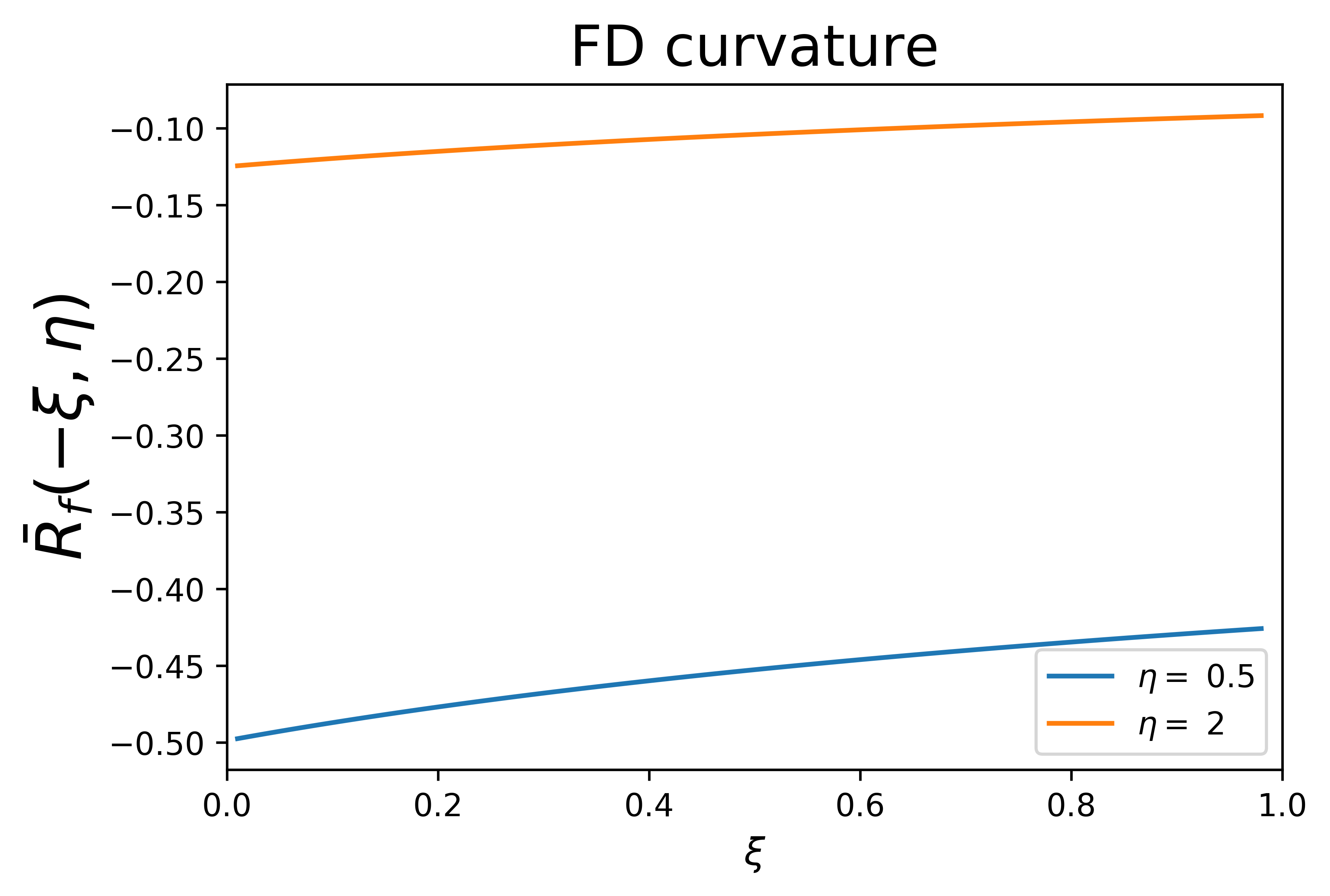}
    \caption{Dimensionless quantity $\bar{R}_f$, related to metric determinant as \eqref{curvaturefermi} for fermions.}
    \label{fig:BoAfermions}
  \end{minipage}
\end{figure}

\begin{equation} \label{metricFD}
    \begin{split}
        g_{11} = - \pdv{^2 F}{\lambda^1 \partial \lambda^1} = -\pdv{U}{\lambda^1} =& - \kappa \frac{\Gamma(\eta+3)}{\beta^{\eta+3}} \Li(-\xi,\eta+2) \ , \\
        g_{12} = g_{21} = - \pdv{^2 F}{\lambda^2 \partial \lambda^1} = -\pdv{N}{\lambda^1} =& - \kappa \frac{\Gamma(\eta+2)}{\beta^{\eta+2}} \Li(-\xi,\eta+1) \ , \\
        g_{22} = - \pdv{^2 F}{\lambda^2 \partial \lambda^2}  = -\pdv{N}{\lambda^2} =& - \kappa \frac{\Gamma(\eta+1)}{\beta^{\eta+1}} \Li(-\xi,\eta) \ . \\
    \end{split}
\end{equation}
The derivatives were taken using \eqref{Polylogderivative} and \eqref{xianddev}. The metric determinant can be directly calculated from \eqref{metricFD},

\begin{equation} \label{detFD}
    g \doteq \left( \frac{\kappa}{\beta^{\eta+2}} \right)^2 \bar{g}_f \ , \quad \text{where} \quad \bar{g}_f= \mathcal{A}(-\xi,\eta)  \ ,
\end{equation}
and $\mathcal A(x,\eta)$ is the unitless quantity defined as

\begin{equation}\label{adef}
\mathcal A(x,\eta) \doteq 
\det \left[\begin{matrix}
{\Gamma(\eta+3)} \Li( x,\eta+2) & {\Gamma(\eta+2)}\Li( x,\eta+1) \\
{\Gamma(\eta+2)} \Li( x,\eta+1) &  {\Gamma(\eta+1)} \Li( x,\eta)
\end{matrix}\right] \ ,
\end{equation}
a graph for $ \bar{g}_f$ is presented in Fig. \ref{fig:Afermions}.
Similarly, the scalar curvature can be calculated from \eqref{curvature2d} obtaining

 \begin{equation}\label{curvaturefermi}
     R = \frac{1}{2g^2} \frac{\kappa^3}{\beta^{3\eta+7}} \mathcal B(-\xi,\eta) =  \frac{\beta^{\eta+1}}{2\kappa} \bar{R}_f  \ ,
 \end{equation}
 where
\begin{equation}
     \bar{R}_f \doteq \frac{\mathcal B(-\xi,\eta)  }{\mathcal A(-\xi,\eta)^2}    \ ,
\end{equation}
and $\mathcal B(x,\eta)$ is defined as
\begin{equation}\label{bdef}
\mathcal B(x,\eta) \doteq 
\det \left[ \begin{matrix}
{\Gamma(\eta+3)} \Li( x,\eta+2) & {\Gamma(\eta+2)} \Li( x,\eta+1) & {\Gamma(\eta+1)} \Li( x,\eta) \\ 
{\Gamma(\eta+4)} \Li( x,\eta+2) & {\Gamma(\eta+3)} \Li( x,\eta+1) & {\Gamma(\eta+2)} \Li( x,\eta) \\ 
{\Gamma(\eta+3)} \Li( x,\eta+1) & {\Gamma(\eta+2)} \Li( x,\eta) & {\Gamma(\eta+1)} \Li( x,\eta-1)
\end{matrix} \right] \ .
\end{equation}
As discussed before, $\kappa$ has units so that $\frac{\beta^{\eta+1}}{\kappa}$ is unitless. A graph for $R_f$ in \eqref{curvaturefermi} is presented in Fig. \ref{fig:BoAfermions}. Also a visual representation of $R$ in  terms of $\beta$ and $\xi$ is presented in Fig. 3. These graphs are simple and smooth, meaning that for FD models there are no anomalies or singularities. Furthermore, it also can be seen from the Fig. 2 and Fig. 3 that the curvature for FD gases is negative.

\begin{figure}%[h]
\label{fig:Rfermions}
  \begin{minipage}[b]{0.5\textwidth}
    \includegraphics[width=\textwidth]{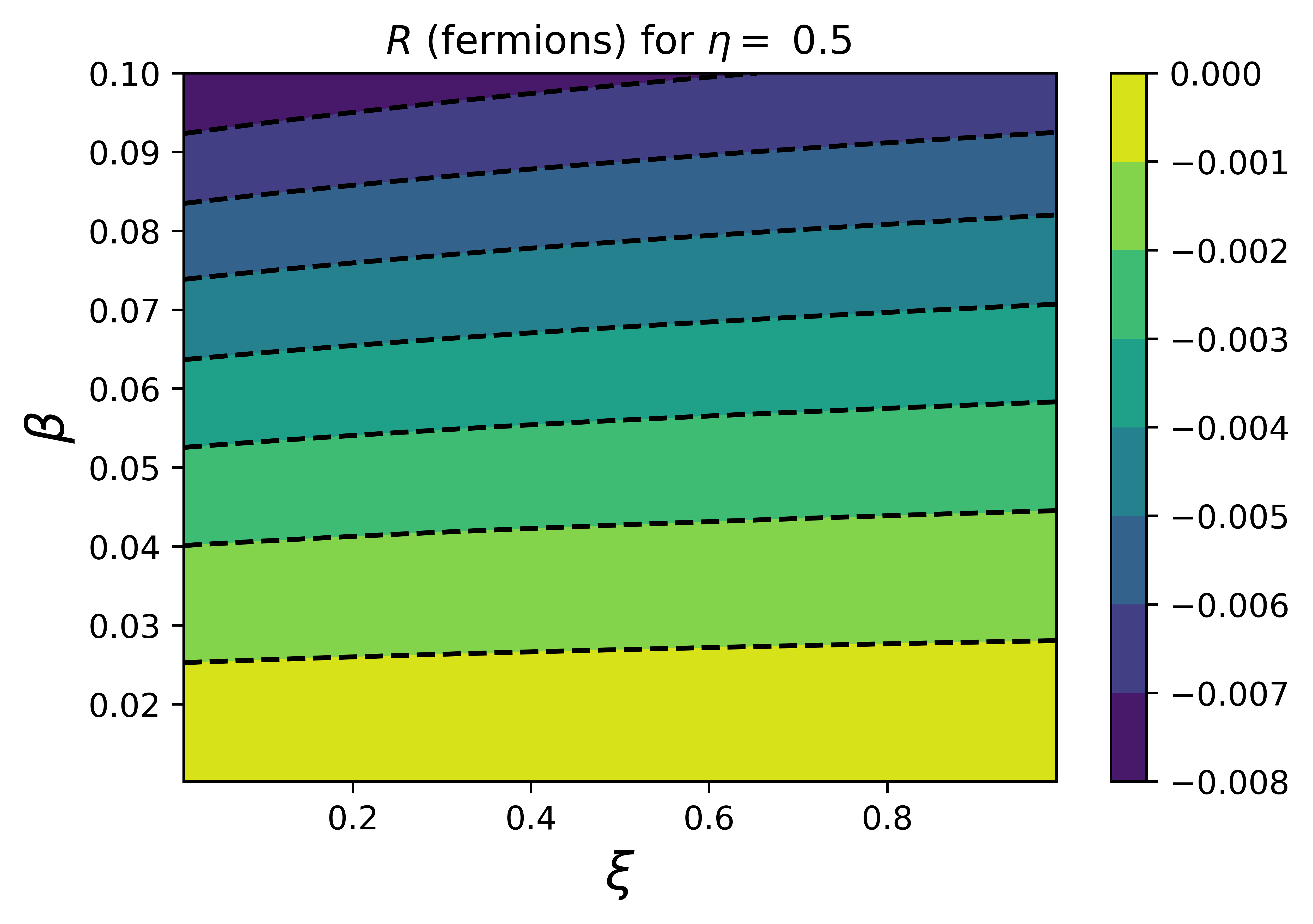}
  \end{minipage}
  \hfill
  \begin{minipage}[b]{0.5\textwidth}
    \includegraphics[width=\textwidth]{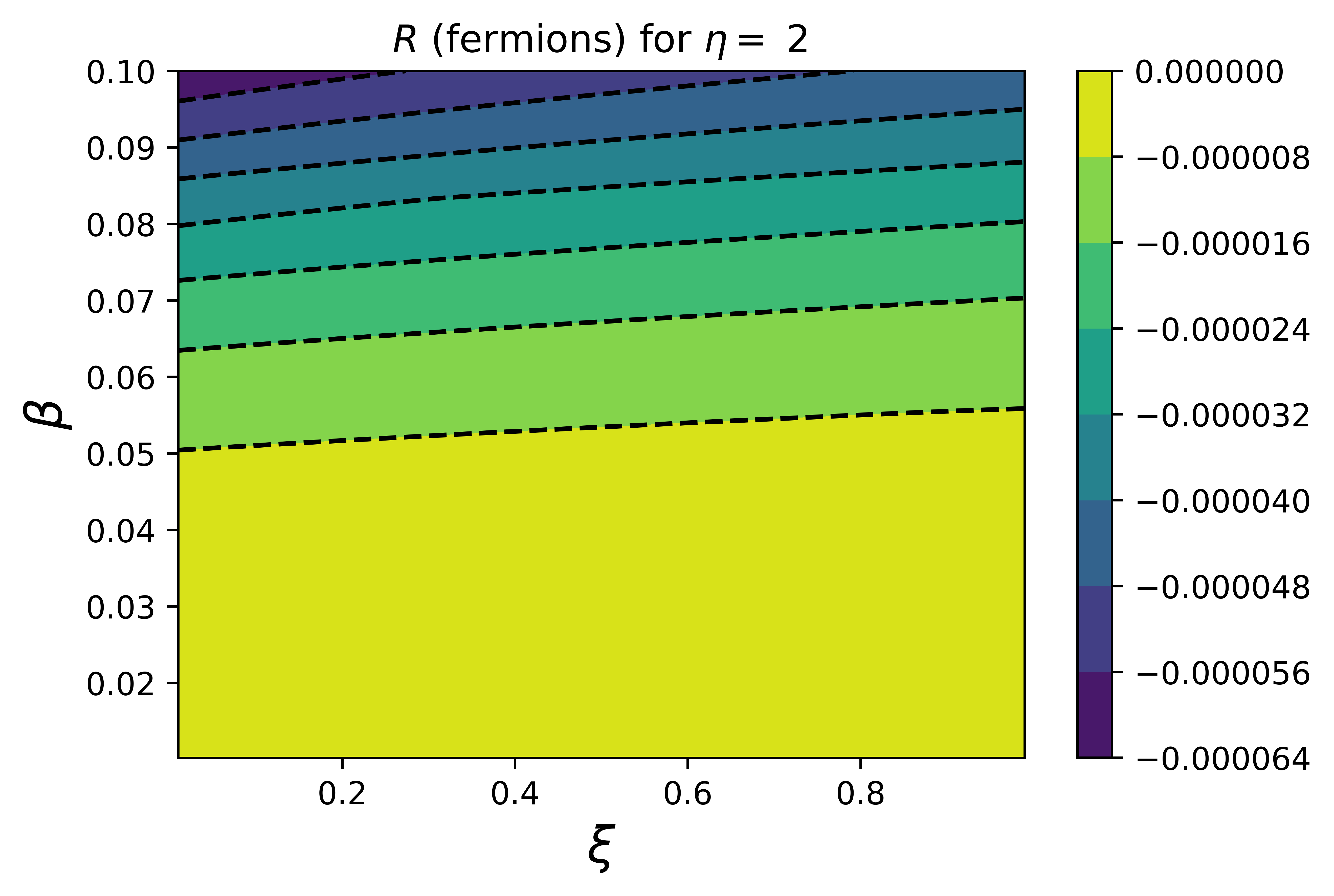}
  \end{minipage}
 
    \caption{Countour plot of $R$ in \eqref{curvaturefermi} as a function of $\xi$ and $\beta$  -- in units of $\kappa=1$ -- for fermions in the values of $\eta = \half 1$ and $\eta = 2$. }
\end{figure}

\section{Information geometry - Bosons}
\paragraph{}
A similar geometric description of bosonic gases, meaning obtain FRIM from differentiating \eqref{Fcon} as it was done in the previous section for fermions, would yield a metric determinant and scalar curvature given, respectively, as

    \begin{equation}\label{gandR}
    g_0 =\left( \frac{\kappa}{\beta^{\eta+2}} \right)^2 \mathcal A(\xi,\eta)
    \quad \text{and} \quad
     R_0 = - \frac{\beta^{\eta+1}}{2\kappa} \frac{\mathcal B(\xi,\eta)}{(\mathcal A(\xi,\eta))^2} \ ,
     \end{equation}
where $\mathcal A$ and $\mathcal B$ are the same as defined in \eqref{adef} and \eqref{bdef}.
These are denoted as $g_0$ and $R_0$ to differentiate as they do not take into account the particles in the ground state. Works such as \cite{Janyszek90, Oshima99} are limited to this description. In this section, we will calculate FRIM and other geometrical quantities appropriately taking into account the ground state.

The ground state term correction is necessary as the continuous approximation \eqref{density} assigns no particles in the ground state, $G(0) = 0 $. An accurate description needs to account for condensation as \eqref{Averages4}, and consequentially \eqref{gandR} would be an approximation for high temperatures. This is achieved by an ad-hoc addition to the term corresponding to the ground state, $\epsilon = 0$, in \eqref{Averages3}:

\begin{equation}
    N_0 = \frac{1}{e^{\lambda^2}-1} \ .
\end{equation}
So that $N$ in \eqref{Averages4} for bosons becomes
\begin{equation}\label{NBE}
    N = A_2 =  \kappa \frac{ \Gamma(\eta+1)}{\beta^{\eta+1}} \Li( \xi,\eta+1)   + \frac{1}{\xi^{-1}-1} \ .
\end{equation}
and no change in the internal energy $U$ is necessary as the ground state does not contribute to the average energy.

In this correction, since all we have are the final expressions for $U$ in \eqref{Averages4} and $N$ in \eqref{NBE}, we will proceed with the metric calculation using  \eqref{gequalsC} and \eqref{corrmetric}, this is akin to the thermodynamical geometry of Ruppeiner \cite{Ruppeiner79}. This leads to the metric terms 

\begin{equation} \label{metricBE}
    \begin{split}
        g_{11} =  -\pdv{U}{\lambda^1} =& \kappa \frac{\Gamma(\eta+3)}{\beta^{\eta+3}} \Li(\xi,\eta+2) \ , \\
        g_{12} = g_{21} =  -\pdv{N}{\lambda^1} =&  \kappa \frac{\Gamma(\eta+2)}{\beta^{\eta+2}} \Li(\xi,\eta+1) \ , \\
        g_{22} = -\pdv{N}{\lambda^2} = &  \kappa \frac{\Gamma(\eta+1)}{\beta^{\eta+1}} \Li(\xi,\eta) + \frac{\xi}{(1-\xi)^2} \ ,
    \end{split}
\end{equation}
where the only metric term that is influenced by the particles in the ground state is $g_{22}$.

Then, the metric determinant takes the form

\begin{equation}\label{gbosons}
    g = \left(\frac{\kappa}{\beta^{\eta+2}} \right)^2 \bar{g}_b(\xi,\eta) \ , \quad \text{where} \quad \bar{g}_b(\xi,\eta) \doteq \mathcal A(\xi,\eta) + 
    \frac{\beta^{\eta+1}}{\kappa}  \mathcal A_c(\xi,\eta) \ ,
\end{equation}
where $\bar{g}_b$ is dimensionless, $\mathcal A$ is the same as in \eqref{adef} and $\mathcal A_c(x,\eta)$ is defined as

\begin{equation}\label{acdef}
    \mathcal A_c(x,\eta) \doteq 
    \det \left[\begin{matrix}
    {\Gamma(\eta+3)} \Li( x,\eta+2) &0 \\
    {\Gamma(\eta+2)} \Li( x,\eta+1) &  \frac{x}{(1-x)^2}
    \end{matrix}\right] \ .
\end{equation}
Moreover, if the $\mathcal A_c$ term is ignored in \eqref{gbosons} $g$ reduces to $g_0$ in \eqref{gandR}.
A graph for $\bar{g}_b$ is presented in Fig. 4.
We can see that the ground state term makes so that $g$ grows faster than $g_0$  with a rate of change that is increasing with $\beta$. Also, in the graph it can be seen that for $\eta = \half1$, $g_0$ diverges for $\xi \rightarrow 1$  while it converges for $\eta=2$. In both cases, however, $g$ diverges for positive values of $\beta$.

\begin{figure}%[h]
  \begin{minipage}[b]{0.45\textwidth}
    \includegraphics[width=\textwidth]{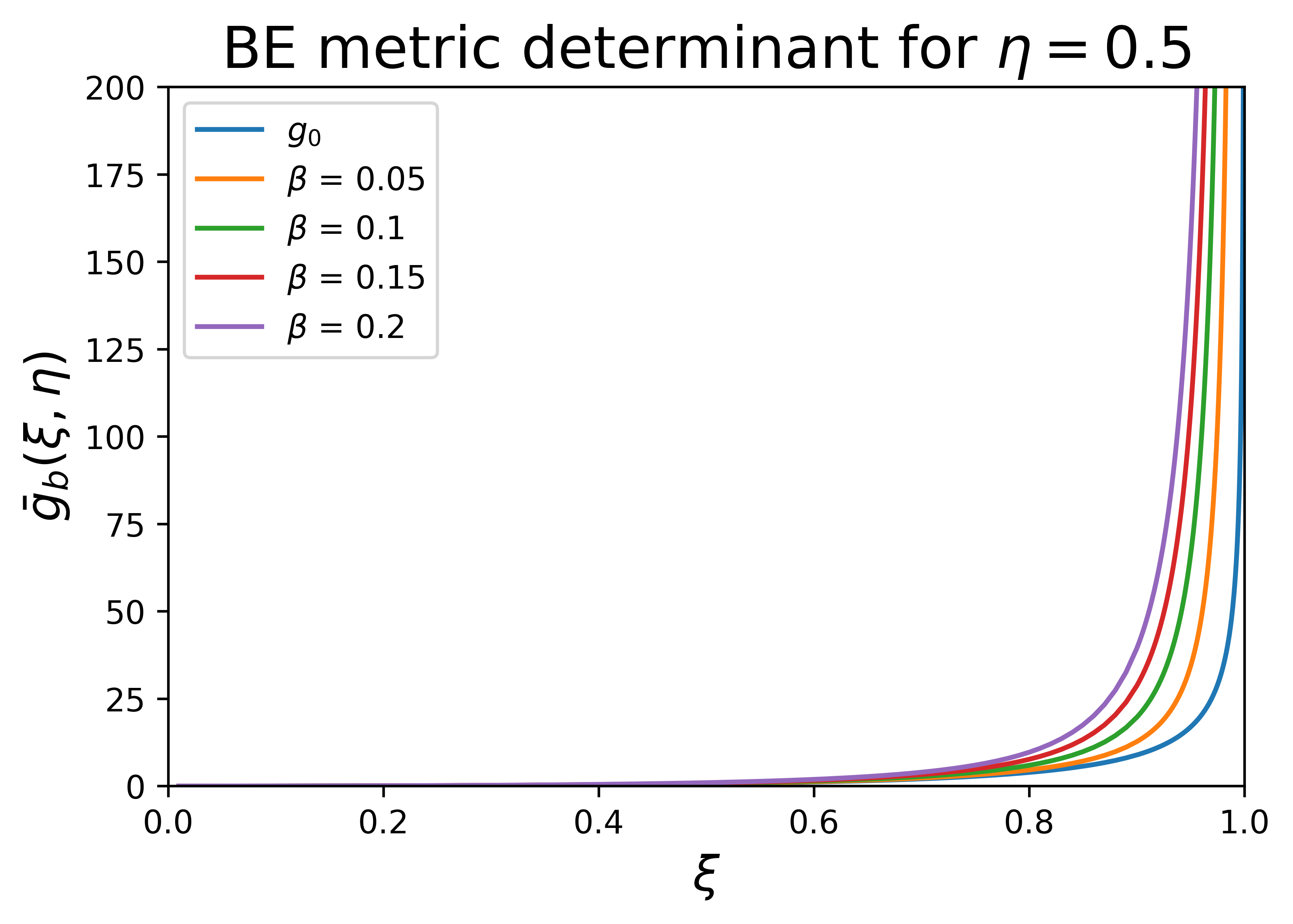}
  \end{minipage}
  \hfill
  \begin{minipage}[b]{0.45\textwidth}
    \includegraphics[width=\textwidth]{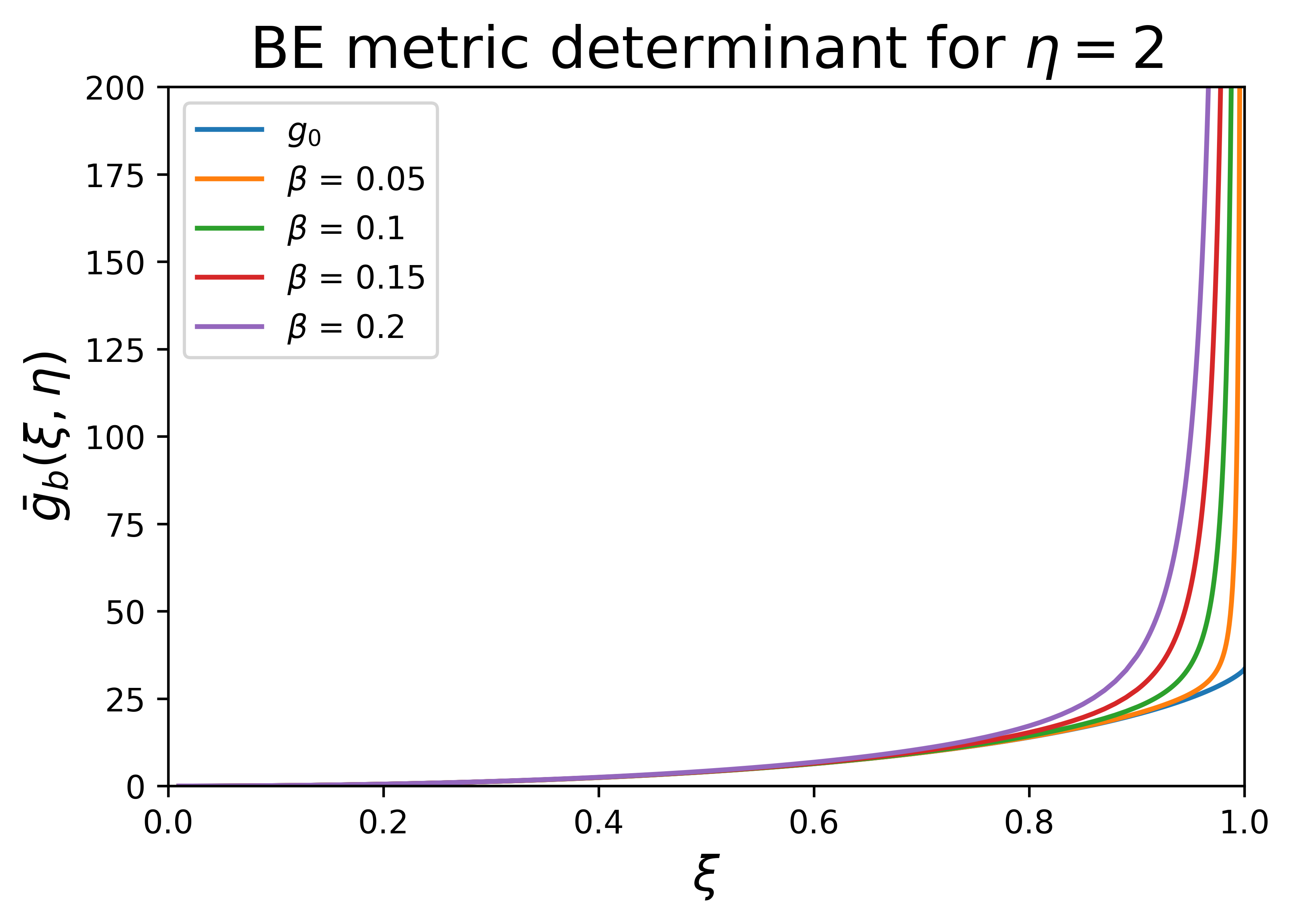}
  \end{minipage}
    \label{fig:Abosons} 
    \caption{Dimensionless quantity $\bar{g}_b$, related to metric determinant as \eqref{gbosons}  for different values of $\beta$ (in units of $\kappa=1$) for bosons in the values of $\eta = \half 1$ and $\eta = 2$. }
\end{figure}

\begin{figure}%[h]
  \begin{minipage}[b]{0.45\textwidth}
    \includegraphics[width=\textwidth]{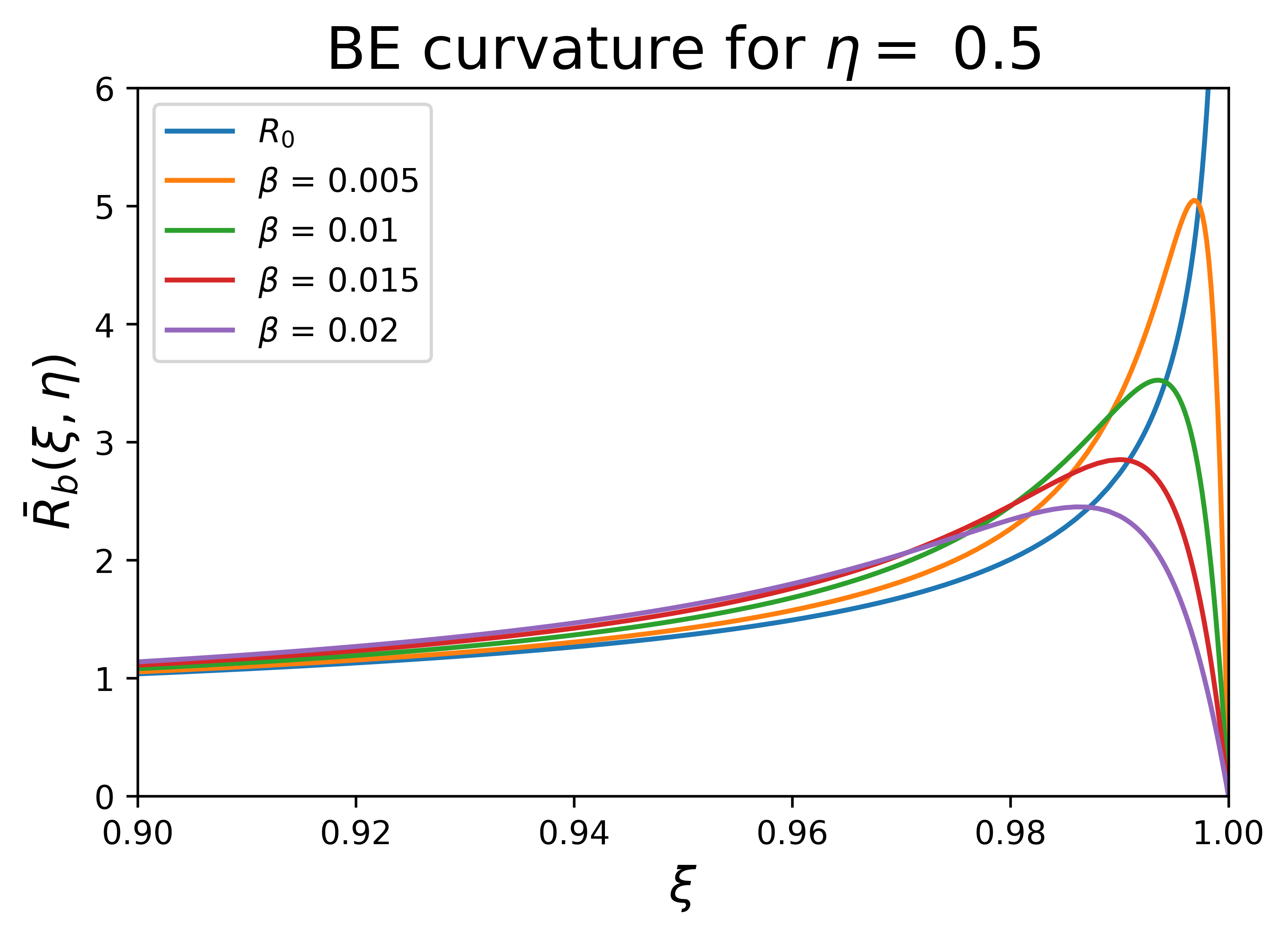}
  \end{minipage}
  \hfill
  \begin{minipage}[b]{0.45\textwidth}
    \includegraphics[width=\textwidth]{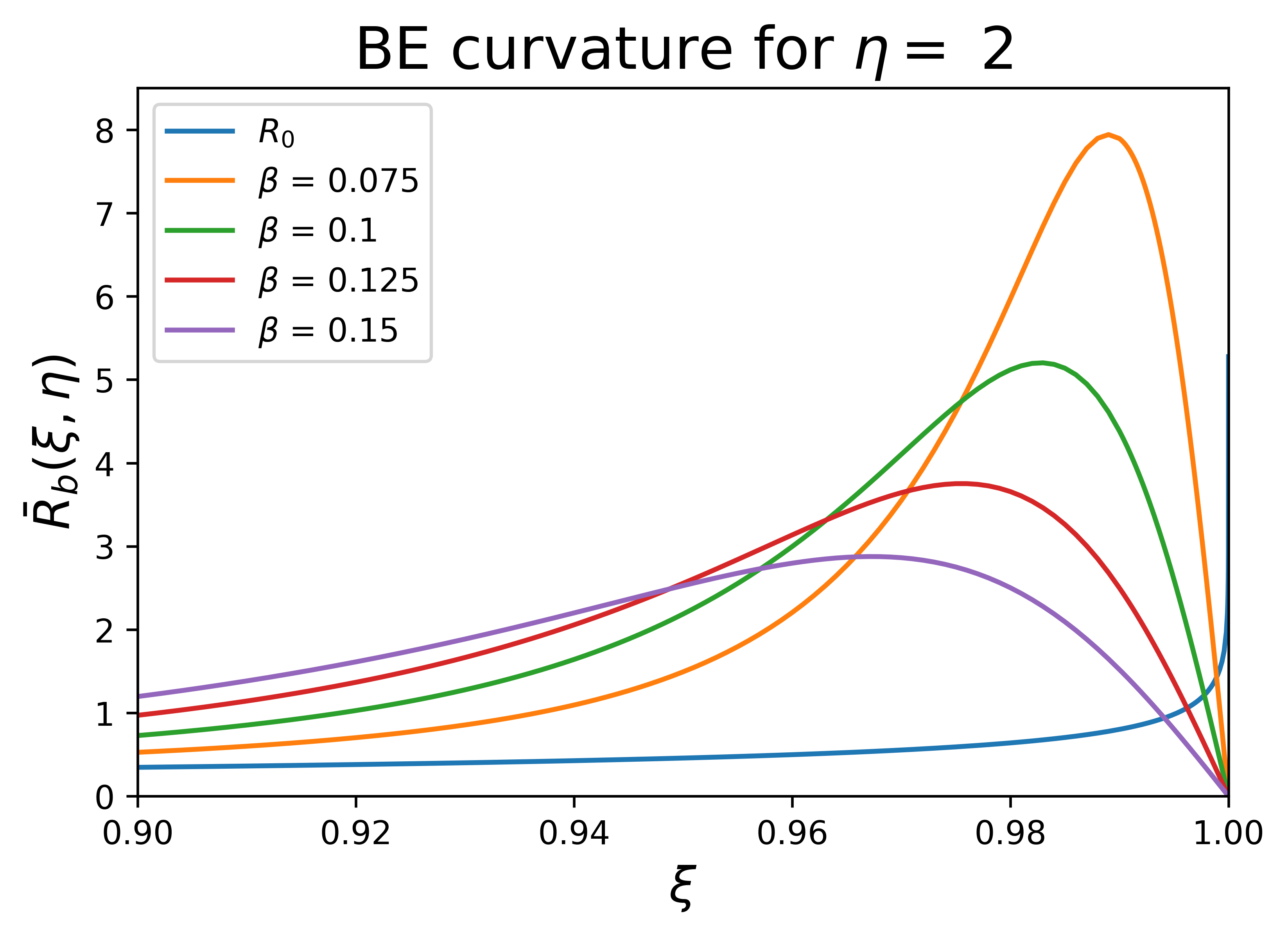}
  \end{minipage}
    \label{fig:Bbosons}
    \caption{ Dimensionless quantity $\bar{R}_b$, related to curvature as \eqref{curvaturebose}  for different values of $\beta$ (in units of $\kappa=1$) for bosons in the values of $\eta = \half 1$ and $\eta = 2$. }
\end{figure}

The scalar curvature can be calculated from \eqref{curvature2d} so that for bosons we have

 \begin{equation}\label{curvaturebose}
     R = - \frac{1}{2g^2} 
     \left[ \frac{\kappa^3}{\beta^{3\eta+7}}\mathcal B(\xi,\eta)   + \frac{\kappa^2}{\beta^{2\eta+6}} \mathcal B_c(\xi,\eta)       \right]
    = - \frac{\beta^{\eta+1}}{2\kappa}  \bar{R}_b(\xi,\eta) \ ,
     %\left[      \frac{B(\xi,\eta) + \frac{\beta^{\eta+1}}{\kappa} B_c(\xi,\eta) }{(A(\xi,\eta)+  \frac{\beta^{\eta+1}}{\kappa}A_c(\xi,\eta) )^2}      \right] \ ,
\end{equation}
where
\begin{equation}\label{curvaturebose2}
     \bar{R}_b(\xi,\eta) \doteq \frac{\mathcal B(\xi,\eta) + \frac{\beta^{\eta+1}}{\kappa} \mathcal B_c(\xi,\eta) }{(\mathcal A(\xi,\eta)+  \frac{\beta^{\eta+1}}{\kappa} \mathcal A_c(\xi,\eta) )^2}    
\end{equation}
$\mathcal B$ is the same as in \eqref{bdef} and $\mathcal B_c(x,\eta)$ is defined as

\begin{equation} \label{bcdef}
\mathcal B_c(x,\eta) \doteq 
\det \left[ \begin{matrix}
{\Gamma(\eta+3)} \Li( x,\eta+2) & {\Gamma(\eta+2)} \Li( x,\eta+1) &\frac{x}{(1-x)^2} \\ 
{\Gamma(\eta+4)} \Li( x,\eta+2) & {\Gamma(\eta+3)} \Li( x,\eta+1) &0 \\ 
{\Gamma(\eta+3)} \Li( x,\eta+1) & {\Gamma(\eta+2)} \Li( x,\eta) &\frac{x(x+1)}{(1-x)^3}
\end{matrix} \right] \ .
\end{equation}
Once again, $R$ reduces to $R_0$ when the ground state term is ignored.
A graph for the $\bar{R}_b$  is presented in Fig. \ref{fig:Bbosons}. From it we can see that $R_0$ diverges for $\xi \rightarrow 1$, however the addition of the ground state term eliminates the divergence, moreover the curvature converges to zero at positive values of $\beta$.

A representation of $R$ \eqref{curvaturebose} in comparison to $R_0$ \eqref{gandR} in terms of $\xi$ and $\beta$  is presented in Fig. 6. In accordance to the graphs of Fig. 5, we can see how the ground state term smooths the growth near $\xi=1$, leads to a fast rise and falls to $0$. That is properly explained as the ground state term makes so that the model includes both the condensed and non-condensed phases. That means, the phase transition disappears as both phases are accurately described by the model.
    
Having calculated the scalar curvature for both FD and BE we can compare them to the classical version. This will be done in the following section.

\begin{figure}%[h]
  \begin{minipage}[b]{0.5\textwidth}
    \includegraphics[width=\textwidth]{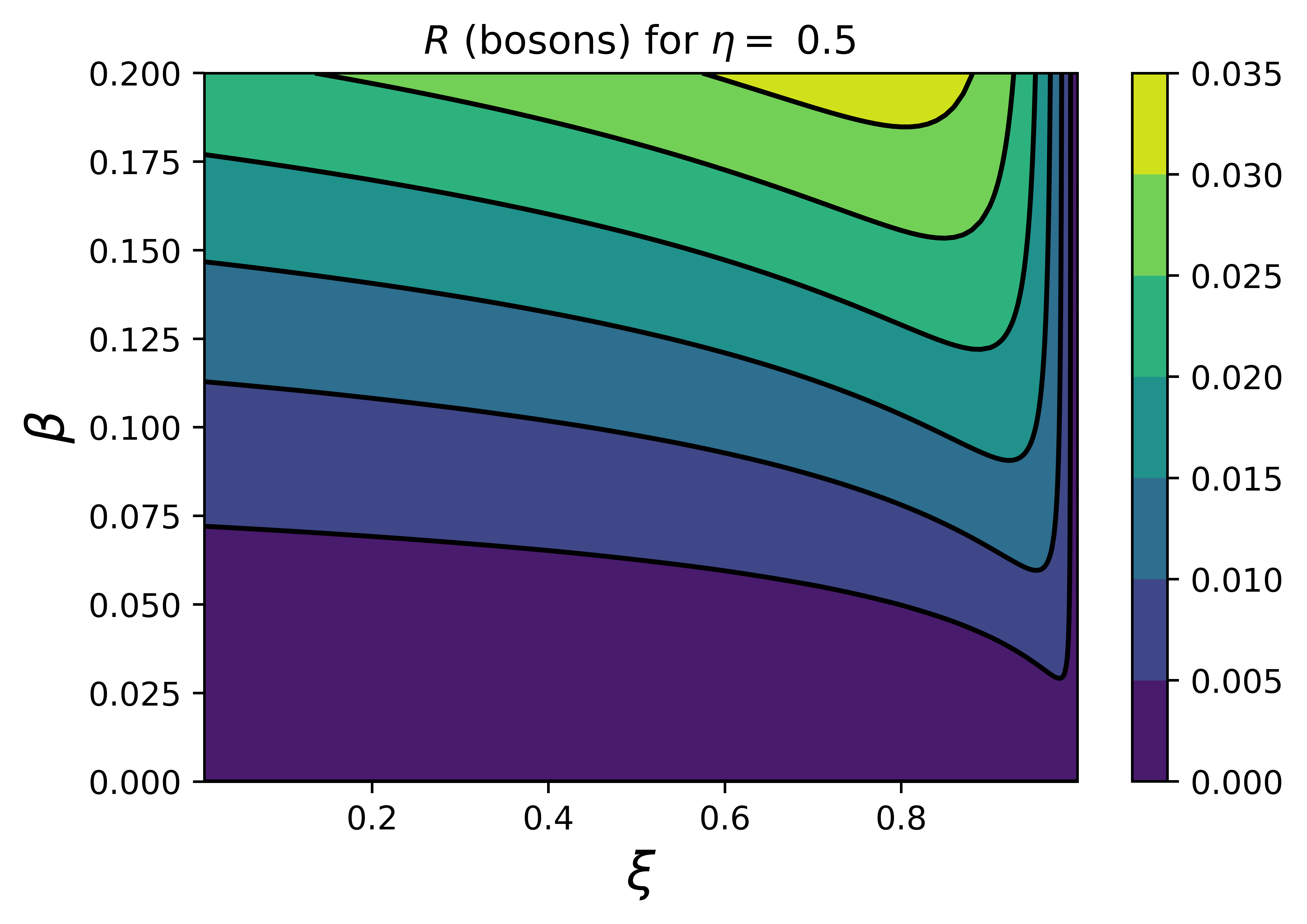}
    \includegraphics[width=\textwidth]{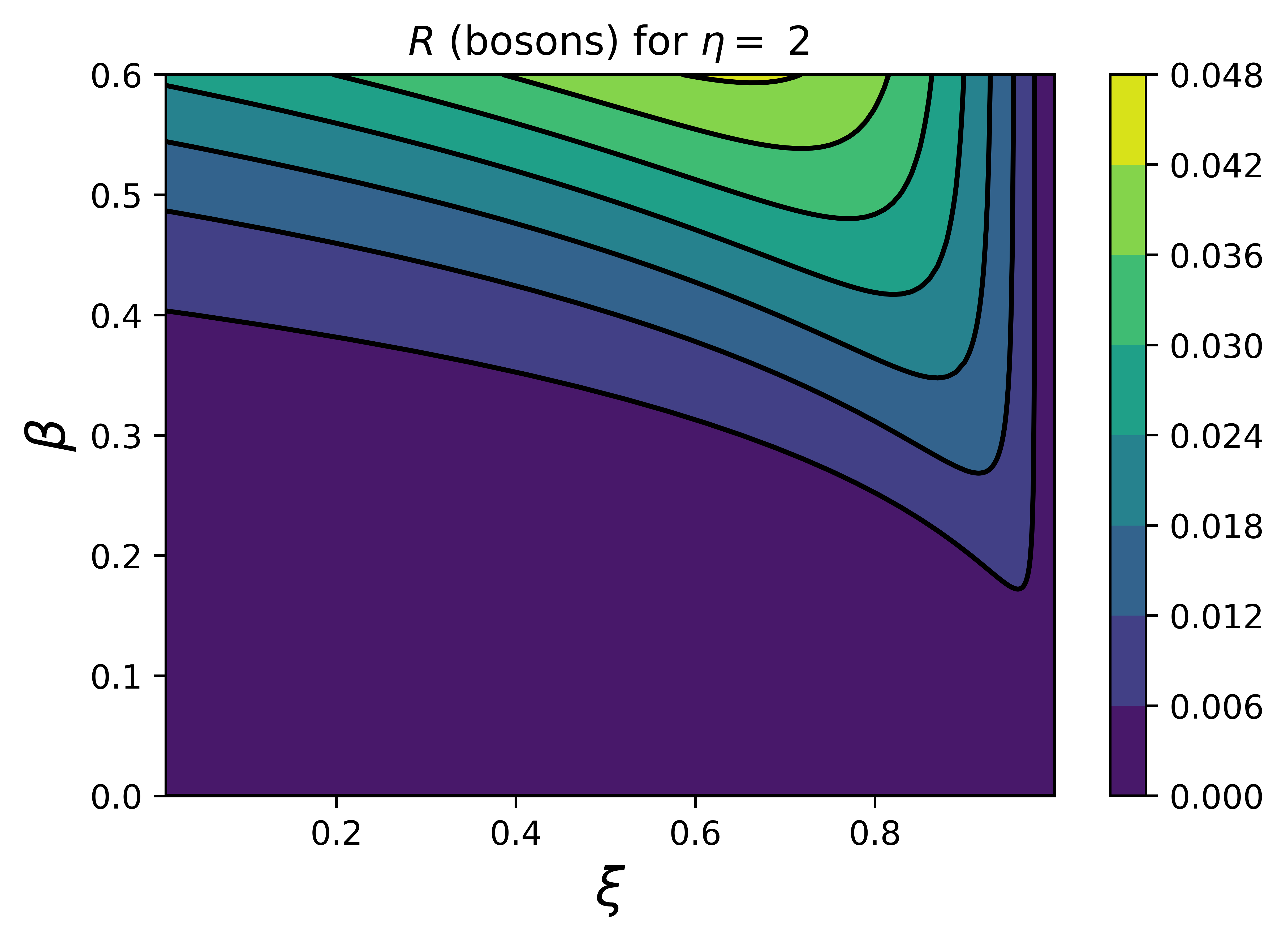}
  \end{minipage}
  \hfill
  \begin{minipage}[b]{0.5\textwidth}
    \includegraphics[width=\textwidth]{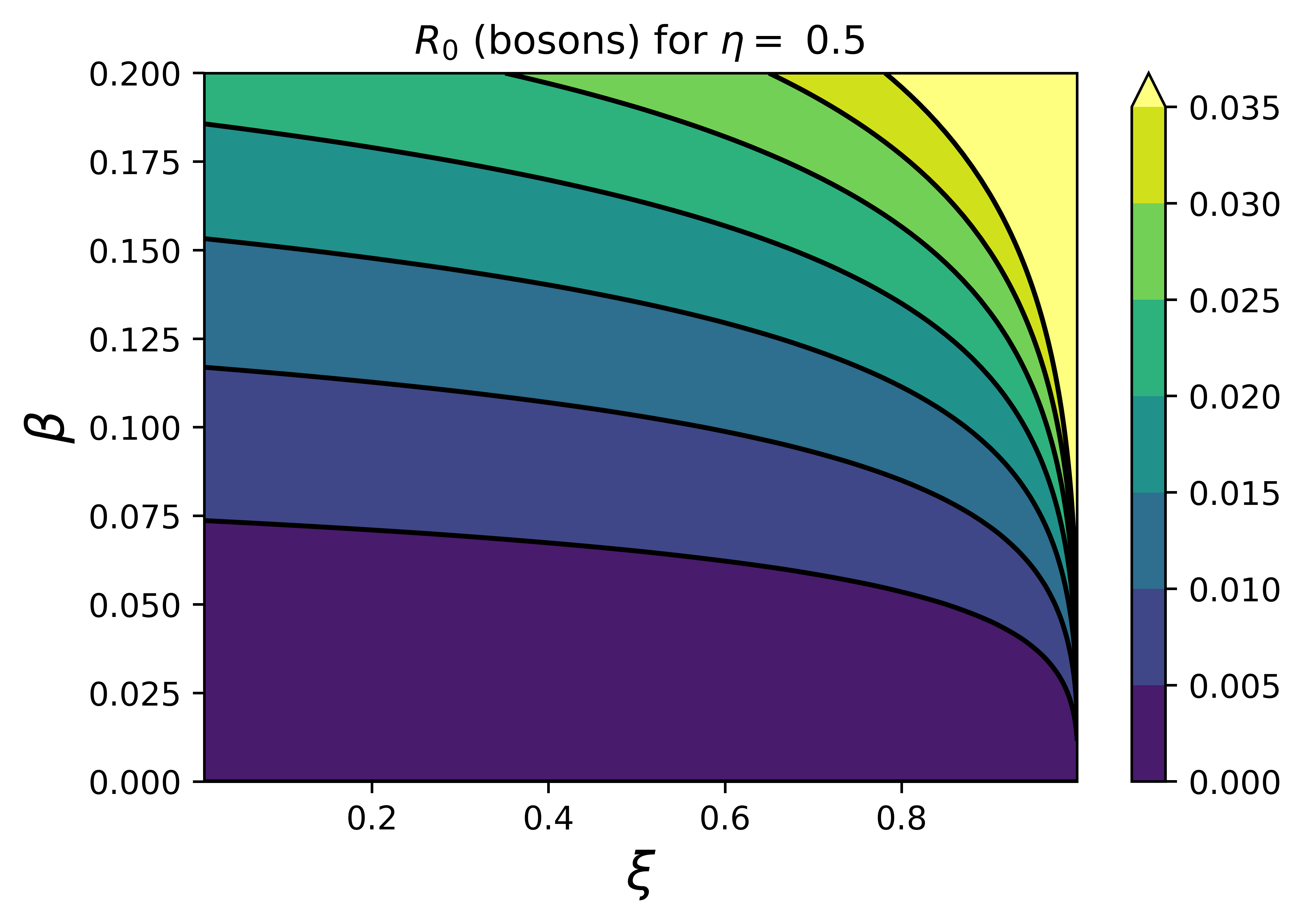}
    \includegraphics[width=\textwidth]{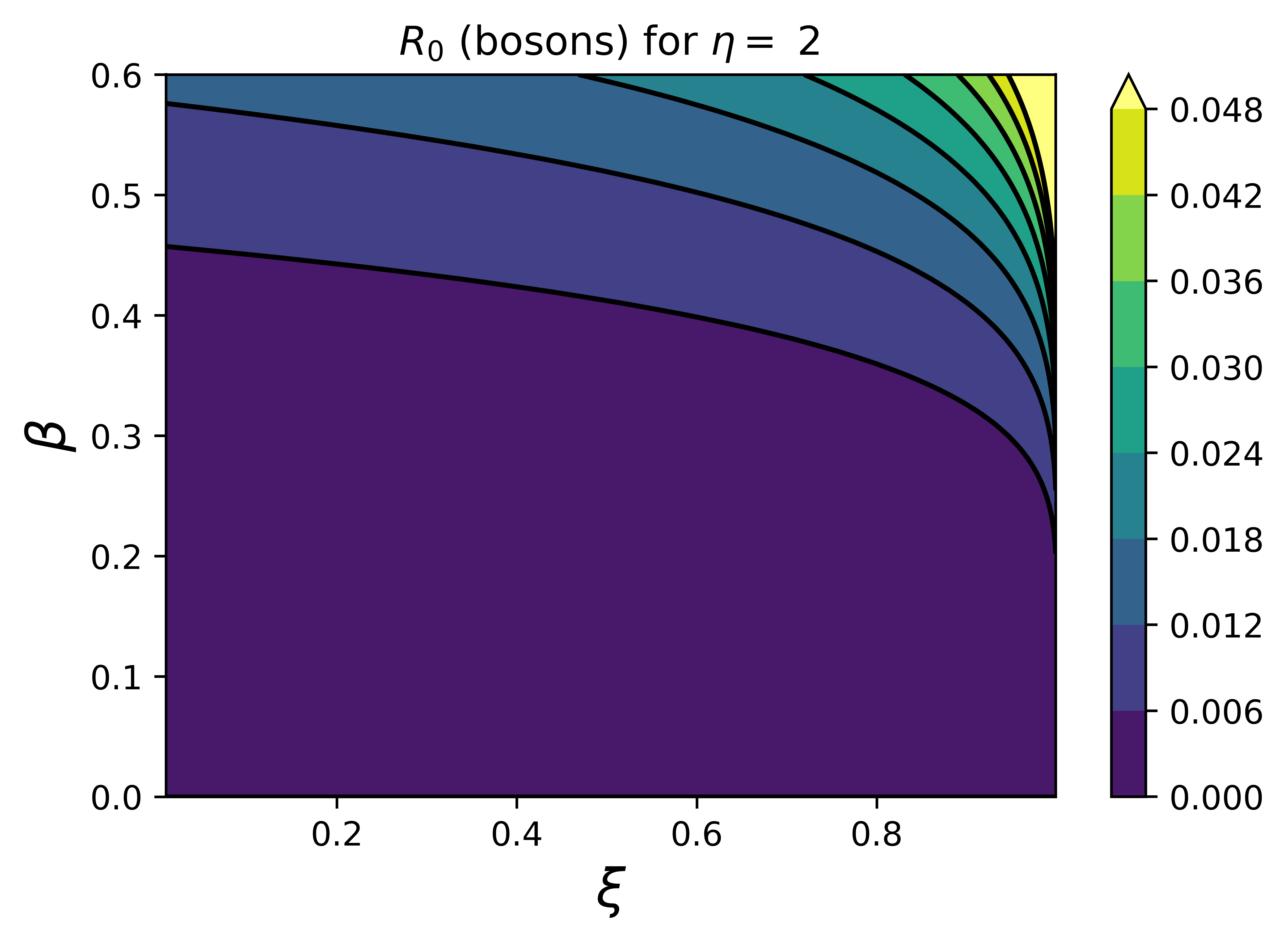}
  \end{minipage}
    \label{fig:conjuctionRandR0} 
    \caption{Countour plot of curvature, $R$ in \eqref{curvaturebose}, compared to their equivalents without the ground state terms, $R_0$ in \eqref{gandR}. Represented as a function of $\xi$ and $\beta$  (units of $\kappa=1$)  in the relevant region for bosons and for the values of $\eta = \half 1$ and $\eta = 2$. }
\end{figure}

\section{Classical limits}
\paragraph{}
For a classical 3-dimensional ideal gas in the grand canonical ensemble the free-energy can be calculated as

\begin{equation}
    F(\lambda) = \log \left[ \sum_{n=0}^\infty e^{-\lambda^2 n} \frac{V^n}{h^{3n} n!}  \prod_{k=1}^n \int d p_k \ {e^{-\lambda^1 \frac{p_k^2}{2m}}} \right]  \ ,
\end{equation}
where $m$ is the mass of particles in the gas. In the thermodynamical limit ($N  \rightarrow \infty$) it can be written in the closed form

\begin{equation}\label{FIG}
      F(\lambda) = - \kappa_{\text{ig}}  \frac{\Gamma(\half3)}{\beta ^{\half{3}}}  \xi \ , \quad \text{where} \quad \kappa_{\text{ig}} = \frac{ V}{4\pi^2}\left( \frac{2m}{\hbar^2}\right)^{\half{3}} \ ,
\end{equation}
which is in accordance with the $\kappa$ calculated for the quantum three dimensional particle in a box (see Table \ref{tab:tablequantum}), here $\xi$ is defined as in \eqref{xianddev}. 
For a comparison between \eqref{FIG} and its equivalent for quantum gases \eqref{Fcon}, it is useful to recall the series expansion for polylogarithms \eqref{Polylogdef}, around $\xi = 0$ meaning

\begin{equation} \label{plfirst}
    \Li(\xi,\varphi) = \xi + o(\xi) \ .
\end{equation}
This implies that the free energy for both FD and BE gases \eqref{Fcon} of a three dimensional particle in a box is equivalent -- meaning has the same dependence to the LM -- to the classical ideal gas in the limit $\xi \rightarrow 0$.

The free energy for the ideal gas leads to a FRIM calculated as

\begin{equation} \label{metricIG}
    \begin{split}
        g_{11} = - \pdv{^2 F}{\lambda^1 \partial \lambda^1} = -\pdv{U}{\lambda^1} =&  \kappa_{\text{ig}} \frac{\Gamma(\half7)}{\beta^{\half7}} \xi \ , \\
        g_{12} = g_{21} = - \pdv{^2 F}{\lambda^2 \partial \lambda^1} = -\pdv{N}{\lambda^1} =&  \kappa_{\text{ig}} \frac{\Gamma(\half5)}{\beta^{\half5}} \xi  \ , \\
        g_{22} = - \pdv{^2 F}{\lambda^2 \partial \lambda^2}  = -\pdv{N}{\lambda^2} =&  \kappa_{\text{ig}} \frac{\Gamma(\half3)}{\beta^{\half3}} \xi \ . \\
    \end{split}
\end{equation}
and this leads to the metric determinant and  curvature

\begin{equation}
    g = \left(\frac{\kappa \xi }{\beta^{\half3}}\right)^2 \Gamma(\half3) \Gamma(\half5) \quad \text{and} \quad R=0 \ .
\end{equation}
Note that although the free energy of the ideal gas \eqref{FIG} reduces to the free energy of FD and BE statistics in the limit $\xi \rightarrow 0$, the curvature for FD \eqref{curvaturefermi} and BE \eqref{curvaturebose} statistics do not converge to zero in the \change{same limit.}
\change{Rather in this regime curvature for FD statistics is given by}

\begin{equation}\label{FDlim}
\change{
    R_{FD} =  \frac{\beta^{\half3}}{2\kappa} 
    \lim_{x\rightarrow0^-}\left[      \frac{\mathcal B(x,\half{1}  ) }{(\mathcal A(x,\half{1}  )) ^2}      \right] \approx - 0.4987 \ \frac{\beta^{\half3}}{2\kappa}  \ .
}
\end{equation}
\change{
the details for the calculation of the limit above are given in Appendix \ref{CurvatureApp}.
Equivalently, the curvature for BE statistics in the same limit, also to be fully calculated in  Appendix  \ref{CurvatureApp}, is
}

\begin{equation}\label{BElim}
\change{
    R_{BE} =  - \frac{\beta^{\half3}}{2\kappa} 
    \lim_{x\rightarrow0^+}\left[     
    \frac{\mathcal B(x,\half1) + \frac{\beta^{\half3}}{\kappa} \mathcal B_c(x,\half1) }{\left(\mathcal A(x,\half1)+  \frac{\beta^{\half3}}{\kappa} \mathcal A_c(x,\half1) \right)^2}   
    \right] \approx 
    \left(\frac{0.6921+5.321 \frac{\beta^{\half3}}{\kappa}}{\left(1.178+3.323\frac{\beta^{\half3}}{2\kappa}\right)^2} \right) \frac{\beta^{\half3}}{ \kappa} \ .
}
\end{equation}
Therefore in order for the curvature in the FD and BE statistics to reduce to the one of the ideal gas it is not sufficient to take the low fugacity limit, rather one should take the high temperature limit ($\beta \rightarrow 0$).

This result gives an interesting insight on the nature of quantum gases.
The limit $\xi \rightarrow 0$, or equivalently  $\lambda^2 = - \frac{\mu}{KT} \rightarrow \infty$, refers to a sparse gas, as per constraints a high value of $\lambda^2$ means a number of particles that is small for the scale of energy.
In other works on the geometric description of quantum gases \cite{Janyszek90,Oshima99} this result is interpreted as the curvature maintains quantum mechanics effects even in the ``classical limit".
We argue against this interpretation by stating that a sparse quantum gas is not equivalent to a classical gas. Instead, one should expect the thermodynamics of a quantum gas to give the same results as in classical mechanics when the energy is bigger than the natural scale of energy of the system given by $\kappa = 1$. Hence $\beta \rightarrow 0$ is a more appropriate classical limit than only $\xi \rightarrow 0$.  Apart from this disagreement, it is important to say that this is a good example of how curvature gives information about the \change{quantum structure of the system} 
while free energy alone would not, stressing the importance and relevance of information geometry in thermodynamics. 

It is also relevant to say that per  \eqref{curvaturebose} the vanishing curvature in the classical limit is fundamentally different from the convergence to zero in BE condensation. In the condensation regime the unitless curvature factor $\bar{R}_b$ converges to zero as the limit of the fugacity $\xi$ approaches one. In the classical limit, instead, $\bar{R}_b$ is multiplied by the reciprocal of temperature $\beta $ approaching zero. %In the interpretation of scalar curvature as a measure of effective interactions, the classical limit would indicate a gas for which quantum effective interactions exist but become less relevant as the temperature increases. For low temperatures, instead, these efective interactions would vanish as the gas approaches condensation.

\section{Conclusion}
\paragraph{}
We obtained closed form expressions for the information metric of FD statistics in the continuous approximation \eqref{metricFD}, they also lead to the metric determinant and volume element  \eqref{detFD} and scalar curvature \eqref{curvaturefermi}. Analogously  we obtain similar expressions for the information metric of BE statistics in \eqref{metricBE} leading to the metric determinant \eqref{gbosons} and curvature \eqref{curvaturebose}.
These results are more general than the ones previously obtained in the literature \cite{Janyszek90,Oshima99,Mirza10,Quevedo15} as they are calculated for a generic density of states exponent $\eta$, which opens for a IG description of several quantum mechanics models as illustrated in  Table \ref{tab:tablequantum}.
Also it is the first time to the best of our knowledge that the ground state correction is being applied in these calculations. Therefore our analysis is carried out with the appropriate information metric for a system evolving into BE condensation. 

As it can be seen from Fig. 5 and Fig. 6, the qualitative behavior of curvature in the BE ideal gas is deeply altered by the addition of the ground state term. In the limit of fugacity approaching unity, the scalar curvature converges to zero  instead of diverging. This result presents a directly calculated counterexample to  the hypothesis of curvature always diverging along a phase transition as BE condensation happens in that limit. \change{This also challenges the interpretation for which the curvature measures microscopic interactions as the condensation is where the effective attraction of bosons is stronger. }

On that sense, we observe that  \eqref{NBE} gives an accurate description of a system in which some particles are in the ground state and others are in the excited states, therefore the curvature does not diverge in a model that correctly describes both phases. In that understanding,  our result suggests  that the conjecture might  be restated as: when the thermodynamical model only describes a single phase, a divergence in scalar curvature indicates where the probability model \textquotedblleft breaks down\textquotedblright, which is usually identified as a phase transition.

The peculiarity of our proposed information geometric characterization of an
ideal bosonic gas that condensates resides in the removal of the singular
behavior of the scalar curvature of a manifold where regions with
structurally different phases are no longer present. The absence of these
distinct phases, in turn, is a consequence of having taken into account in
an appropriate manner the ground state of the ideal bosonic gas. Despite its
originality and broader applicability, our information geometric analysis is
limited to non-interacting gases, further study is necessary on the physical meaning of the scalar curvature in information geometry. However, building on the work presented here, we are confident we
will keep improving our comprehension of these fascinating quantum
mechanical phenomena in future scientific efforts.

\section*{Acknowledgments}
\paragraph{} We would like to thank A. Caticha, F. Xavier Costa and D. Robbins for important discussions during the development of the present article.

P.P. was financed in part by  CNPq -- Conselho Nacional de Desenvolvimento Científico e Tecnológico-- (scholarship GDE 249934/2013-2)
\section*{Appendix}
\appendix

\section{Proof of the identities in Sub-Section \ref{IG}}\label{IGProofs}
\paragraph{}
In this appendix we will prove \emph{Identities 1-4} presented in Sub-Section \ref{IG} and whose results were used to calculate the metric and scalar curvature throughout the paper. From the FRIM definition  \eqref{firstmetric} we can calculate the metric terms for Gibbs distributions \eqref{canonicaldefinition} using  
\begin{equation}
\label{logderivatives}
    \frac{\partial\log \rho(x|A)}{\partial \lambda^\nu}  =  \frac{\partial }{\partial \lambda^\nu}  \left(-  \lambda^\mu a_\mu(x) \right)  - \frac{\partial \log Z }{\partial \lambda^\nu}  =  A_\nu-a_\nu(x)  \ .
\end{equation}
So that \eqref{firstmetric} becomes
\begin{equation} \label{metriciscovariance}
      g_{\mu\nu} =\langle (A_\mu-a_\mu(x)) (A_\nu-a_\nu(x))  \rangle = C_{\mu\nu} \ ,
\end{equation} 
proving, therefore \emph{Identity 1}.
If we use this result when lowering the index of the infinitesimal vectors $d\lambda^{\nu}$ we obtain
\begin{equation}\label{lowerindexes}
d\lambda_\mu = g_{\mu\nu} d\lambda^\nu = - \pdv{A_\mu}{\lambda^\nu} d\lambda^\nu = - dA_\mu \ ,
\end{equation}
where we used the \eqref{corrmetric} proving \emph{Identity 2}. Also \eqref{lowerindexes} shows that the geometric description in terms of expected values is the same as the one using LM, $g_{\mu\nu} dA^\mu dA^\nu = g_{\mu\nu} d\lambda^\mu d\lambda^\nu$. 
Also, if we substitute  \eqref{Averages0} and  \eqref{partial A} into \eqref{corrmetric} we obtain:
 \begin{equation} %\label{metricderiv}
     g^{\mu\nu} = C^{\mu\nu} =- \pdv{}{A_\mu} \pdv{S }{ A_\nu} \quad \text{and} \quad  g_{\mu\nu} = C_{\mu\nu}= -\pdv{}{\lambda^\mu}\pdv{F }{ \lambda^\nu} \ ,
 \end{equation}
 which implies \emph{Identity 3}.
 
Having the metric and inverse metric we can directly calculate the Christoffel coefficients
\begin{equation}
\Gamma_{\sigma\mu\nu} = \frac{1}{2} (\partial_\mu g_{\sigma\nu} +\partial_\nu g_{\mu\sigma} - \partial_\sigma g_{\mu\nu}) = \frac{1}{2} \partial_\sigma g_{\mu\nu} \ ,
\end{equation}
where $\partial_\sigma g_{\mu\nu} = \pdv{}{\lambda^\sigma} g_{\mu\nu}$ and the last equality follows as \emph{Identity 3} implies $\partial_\sigma g_{\mu\nu} = \partial_\nu g_{\mu\sigma}$. This leads to the Riemann curvature tensor
\begin{equation}
    R_{\mu\nu\sigma\upsilon } = \frac{1}{4} g^{o\varpi} \det \left[ \begin{matrix}
\partial_o g_{\mu\upsilon}  & \partial_o g_{\mu\sigma} \\ 
\partial_\varpi g_{\nu\upsilon}& \partial_\varpi g_{\nu\sigma} \\ 
\end{matrix} \right]  \ ,
    %R_{\mu\nu\sigma\upsilon } = \frac{1}{4} g^{o\varpi} (\partial_o g_{\mu\upsilon} \  \partial_\varpi g_{\nu\sigma} - \partial_o g_{\mu\sigma} \ \partial_\varpi g_{\nu\upsilon}  ) \ ,
\end{equation}
due to the symmetries in the the tensor --  anti-symmetric in the first two and second two indexes -- the only non-vanishing components in a two-dimensional manifold are $R_{1212} = R_{2121} = -R_{2112} = -R_{1221}$.  The scalar curvature, then, is given by

\begin{equation}
    R = g^{\nu\upsilon} g^{\mu \sigma}  R_{\mu\nu\sigma\upsilon } \ .
\end{equation}
which in two dimensions and using \emph{Identity 3}  is equivalent to 
\begin{equation} \label{Rappendix}
    R = (g^{11}g^{22}-g^{12}g^{21})  R_{1212} =  \frac{1}{2\det g_{\mu\nu} } 
\det \left[ \begin{matrix}
g^{22} &  g^{12} & g^{11} \\ 
\partial_1 g_{11} & \partial_1 g_{12} & \partial_1 g_{22} \\ 
\partial_2 g_{11} & \partial_2 g_{12} & \partial_2 g_{22} \\ 
\end{matrix} \right] \ .
\end{equation}
If we use the formula for inverse of a two dimensional matrix --  $g^{11} = \frac{g_{22}}{\det g_{\mu\nu}}$, $g^{12} = - \frac{g_{21}}{\det g_{\mu\nu}}$,  $g^{21} = - \frac{g_{12}}{\det g_{\mu\nu}}$, and $g^{22} = \frac{g_{11}}{\det g_{\mu\nu}}$  -- \eqref{Rappendix} is equivalent to \emph{Identity 4}.

\change{
\section{On the low fugacity limit of curvature}\paragraph{}
\label{CurvatureApp}
In this appendix we will explicitly calculate all steps needed for finding the low fugacity limit for curvature in FD statistics \eqref{FDlim} and BE statistics \eqref{BElim}. 
For the purposes of our investigation we will assume that the value of $\eta$ is so that all of the following quantities: $\Gamma(\eta+1)$, $ \Gamma(\eta+2)$, $ \Gamma(\eta+3)$, and $\Gamma(\eta+4)$   do not diverge. This is observed in all models presented in Table \ref{tab:tablequantum}, furthermore $\eta > -1$ is sufficient condition for the assumption to hold truth.
}

\change{
Using the first term expansion of polylogarithms \eqref{plfirst} in the calculation of $\mathcal{A}$ in \eqref{adef} we obtain
}
\begin{equation}\label{afirstorder}
    \mathcal{A}(x,\eta) = f(\eta) x^2 + o(x^2) \ ,
\end{equation}
\change{
where $f(\eta) \doteq  \Gamma(\eta+3) \Gamma(\eta+1) -  \Gamma(\eta+2)^2 $.
Equivalently for  $\mathcal{A}_c$ in \eqref{acdef} we obtain
}
\begin{equation}
    \mathcal{A}_c(x,\eta) = f_c(\eta) x^2 + o(x^2) \ ,
\end{equation}
\change{
where $f_c(\eta) \doteq \Gamma(\eta+3)$. Also, in the first term expansion of polylogarithms for $\mathcal{B}$ in \eqref{bdef} we obtain 
}
\begin{equation}
    \mathcal{B}(x,\eta) =  o(x^3) \ ,
\end{equation}
\change{ that means, $\mathcal{B}$ vanishes in order $x^3$. This is not surprising we can see in \eqref{bdef} that the first order expansion of polylogarithms turns  $\mathcal{B}$ into the determinant of a matrix with duplicate lines.
However, since we interested in calculating the limits of curvature in \eqref{FDlim} and \eqref{BElim},  $\mathcal{B}$ needs to be compared to  $\mathcal{A}^2$, and by squaring \eqref{afirstorder} we see that the first non-vanishing term of $\mathcal{A}^2$ is of order $x^4$. Hence, we need to compute $\mathcal{B}$ at least up to order $x^4$.}

\change{If we use the second term expansion of the polylogarithm -- meaning expanding \eqref{Polylogdef} as $\Li(x,\phi) = x + 2^{-\phi} x^2 + o(x^2) $ -- we obtain}
\begin{equation}
    \mathcal{B}(x,\eta) =  h(\eta) x^4 + o(x^4) \ ,
\end{equation}
\change{where}
\begin{equation}\begin{split}
     h(\eta) \doteq 
\frac{1}{2^{\eta+1}}  \left[ - \ \Gamma(\eta+1) \Gamma(\eta+2) \Gamma(\eta+4) +  \frac{3}{2} \Gamma(\eta+1) \Gamma(\eta+3) \Gamma(\eta+3)  \right. \ & \\ 
\left. - \ \frac{1}{2} \Gamma(\eta+2) \Gamma(\eta+2) \Gamma(\eta+3) 
\right] & \ . 
\end{split}\end{equation}
\change{Equivalently for  $\mathcal{B}_c$ in \eqref{bcdef} we obtain } 
\begin{equation}
    \mathcal{B}_c(x,\eta) =   h_c(\eta) x^4 + o(x^4) \ ,
\end{equation}
\change{where}
\begin{equation}\begin{split}
     h_c(\eta) \doteq 
\frac{1}{2^{\eta+1}}  \left[ \Gamma(\eta+2) \Gamma(\eta+4) - \frac{1}{2}\Gamma(\eta+3) \Gamma(\eta+3)  \right]  & \\ +\
2\left[\frac{}{} \Gamma(\eta+3) \Gamma(\eta+3) - \Gamma(\eta+2) \Gamma(\eta+4) \right] &
\ .
\end{split}\end{equation}

\change{Hence, the curvature for FD statistics \eqref{curvaturefermi} in the low fugacity limit is given by}
\begin{equation}\label{FDlimapp}
    \lim_{\xi \rightarrow0} R_{FD} =  \frac{\beta^{\eta+1}}{2\kappa} 
    \lim_{x\rightarrow0^-}\left[      \frac{\mathcal B(x,\eta  ) }{(\mathcal A(x,\eta  )) ^2}      \right] = \frac{ h(\eta)}{(f(\eta))^2} \ \frac{\beta^{\eta+1}}{2\kappa}  \ ,
\end{equation}
\change{
and the curvature for BE statistics \eqref{curvaturebose} in the same limit is given by
}
\begin{equation}\label{BElimapp}
\begin{split}
    \lim_{\xi \rightarrow0}  R_{BE} &=  - \frac{\beta^{\eta+1}}{2\kappa}
    \lim_{x\rightarrow0^+}\left[     
    \frac{\mathcal B(\xi,\eta) + \frac{\beta^{\eta+1}}{\kappa} \mathcal B_c(\xi,\eta) }{\left(\mathcal A(\xi,\eta)+  \frac{\beta^{\eta+1}}{\kappa} \mathcal A_c(\xi,\eta) \right)^2}   
    \right] \\
    &=    -\left[ \frac{ h(\eta)+  h_c(\eta) \frac{\beta^{\eta+1}}{\kappa}}{\left(f(\eta)+f_c(\eta) \frac{\beta^{\eta+1}}{\kappa}\right)^2} \right] \frac{\beta^{\eta+1}}{2\kappa}\ .
\end{split}
\end{equation}
\change{for the 3 dimensional gas in a box, $\eta=\half{1}$, we can calculate  
$h  (\half1) \approx -0.6921$, 
$h_c(\half1) \approx -5.321$, 
$f  (\half1) \approx 1.178$, and
$f_c(\half1) \approx 3.323$. 
Substituting these values \eqref{FDlimapp} becomes \eqref{FDlim} and \eqref{BElimapp} becomes \eqref{BElim} completing the calculation.
}

\bibliographystyle{elsarticle-num}
\bibliography{Elsevier-references}

\end{document}